\documentstyle[aps,epsfig,preprint,floats,tighten,eqsecnum]{revtex}
\draft
\newcommand{\be}{\begin{equation}}
\newcommand{\ee}{\end{equation}\noindent}
\newcommand{\bear}{\begin{eqnarray}}
\newcommand{\ear}{\end{eqnarray}\noindent}
\newcommand{\no}{\noindent}

\newcommand{\Det}{{\rm Det}}

\def\Mneg{\!\!\!\!\!\!\!\!\!\!\!\!\!\!\!\!\!\!\!\!}

\def\Eins{\mathord{1\hskip -1.5pt
\vrule width .5pt height 7.75pt depth -.2pt \hskip -1.2pt
\vrule width 2.5pt height .3pt depth -.05pt \hskip 1.5pt}}

\newcommand{\slD}{\raise.15ex\hbox{$/$}\kern-.57em\hbox{$D$}}
\newcommand{\slpartial}{\raise.15ex\hbox{$/$}\kern-.57em\hbox{$\partial$}}
\newcommand{\slG}{{{\dot G}\!\!\!\! \raise.15ex\hbox {/}}}

\def\GBd12{{\dot G}_{B12}}
\def\PITD{{(4\pi T)}^{-{D\over 2}}}
\def\non{\nonumber}
\def\beqn*{\begin{eqnarray*}}
\def\eqn*{\end{eqnarray*}}

\def\square{\kern1pt\vbox{\hrule height 1.2pt\hbox{\vrule width 1.2pt
   \hskip 3pt\vbox{\vskip 6pt}\hskip 3pt\vrule width 0.6pt}
   \hrule height 0.6pt}\kern1pt}
\def\slash#1{#1\!\!\!\raise.15ex\hbox {/}}

\def\rightvac{\mid 0\rangle}
\def\leftvac{\langle 0\mid}
\def\dps{\displaystyle}

\def\half{{1\over 2}}

\def\fourth{{1\over4}}

\def\e{\mbox{e}}

\def\kinb{{1\over 4}\dot x^2}

\def\4piTD{{(4\pi T)}^{-{D\over 2}}}
\def\4piT4{{(4\pi T)}^{-2}}

\def\Tintm4{{\dps\int_{0}^{\infty}}{dT\over T}\,e^{-m^2T}
    {(4\pi T)}^{-2}}
\def\Tintm{{\dps\int_{0}^{\infty}}{dT\over T}\,e^{-m^2T}}
\def\Tint{{\dps\int_{0}^{\infty}}{dT\over T}}

\def\Dx{\dps\int{\cal D}x}
\def\Dy{\dps\int{\cal D}y}
\def\Dpsi{\dps\int{\cal D}\psi}

\def\bbbz{{\mathchoice {\hbox{$\sf\textstyle Z\kern-0.4em Z$}}
{\hbox{$\sf\textstyle Z\kern-0.4em Z$}}
{\hbox{$\sf\scriptstyle Z\kern-0.3em Z$}}
{\hbox{$\sf\scriptscriptstyle Z\kern-0.2em Z$}}}}
\begin{document}
\newcommand{\info}{{\it Axial Calculations II}
by {\rm Dilkes, McKeon and Schubert}} 
\newcommand{\D}{{\cal D}}
\newcommand{\zero}[1]{{#1}_0} 
\newcommand{\tr}{{\rm tr}}
\newcommand{\sdet}{{\rm sdet}}
\newcommand{\str}{{\rm str}}
\newcommand{\viz}{{\em viz} }
\newcommand{\sign}{{\rm sign}}
\renewcommand{\L}{{\mathcal L}}
\renewcommand{\P}{{\mathcal P}}
\newcommand{\Phat}{\hat{\mathcal P}}
\newcommand{\Tr}{{\rm Tr}}
\preprint{\parbox{2in}{
\flushright IASSNS-HEP-98/104\\ 
LAPTH-710/98\\ hep-th/9812213}}
\title{A New Approach to Axial Vector Model Calculations II}
\newcounter{ouraddress}
\author{F.A. Dilkes$^{(\ref{UWO})}%
$\thanks{fad@julian.uwo.ca \hfill $^\dagger$tmleafs@apmaths.uwo.ca
\hfill $^\ddagger$schubert@lapp.in2p3.fr},
D.G.C. McKeon$^{(\ref{UWO})\dagger}$,
Christian Schubert$^{(\ref{Annecy},\ref{ias})\ddagger}$
\\ 
\vspace{0.5cm}
\refstepcounter{ouraddress}
\label{UWO}
$^{(\ref{UWO})}$Department of Applied Mathematics\\
University of Western Ontario\\
London~CANADA, 
N6A 5B7\\
\vspace{0.5cm}
\refstepcounter{ouraddress}
\label{Annecy}
$^{(\ref{Annecy})}$%
Laboratoire d'Annecy-le-Vieux de Physique Th{\'e}orique
LAPTH\\
Chemin de Bellevue, BP 110 \\
F-74941 Annecy-le-Vieux CEDEX, France\\
\refstepcounter{ouraddress}
\vspace{0.5cm}
\label{ias}
$^{(\ref{ias})}$%
School of Natural Sciences, Institute for Advanced Study,\\
Olden Lane, Princeton, NJ 08540, USA\\
\vspace{0.5cm}} 
\maketitle
\begin{abstract}%
\noindent
We further develop the new approach,
proposed in 
part I
, to computing the heat kernel associated 
with a Fermion coupled to vector and axial vector fields.
We first use the path integral
representation obtained 
for the heat kernel trace in a 
vector-axialvector background to derive a
Bern-Kosower type master formula for the
one-loop amplitude with $M$ vectors and $N$
axialvectors, valid in any even spacetime
dimension. For the
massless case we then generalize this approach to the
full off-diagonal heat kernel.
In the $D=4$ case
the $SO(4)$ structure of the
theory can be broken down to $SU(2) \times SU(2)$
by use of the 't Hooft symbols.
Various techniques for explicitly evaluating the spin part of the
path integral are developed and compared.
We also extend the method to external fermions, and to
the inclusion of isospin. 
On the field theory side, we obtain an extension of
the second order formalism for fermion QED to an
abelian vector-axialvector theory.

\end{abstract}

\renewcommand{\thesection}{\arabic{section}} 
\section{Introduction}

It has been known for a long time that first-quantized 
particle path integrals
are sometimes a useful alternative to standard second-quantized
methods in quantum field theory. Introduced by Feynman himself
at the dawn of relativistic quantum field theory
\cite{feyn48}, particle path integral techniques have found
applications to a variety of problems
in quantum field theory,
ranging from QED ~\cite{Fried,baboca} to
anomalies~\cite{Gaume,FriedanWindey,basvan},
meson-nucleon theory ~\cite{rossch},
and non-perturbative QCD ~\cite{hajese,kksw}.
Only in recent years, however, was this approach also
seriously considered
as a competitor to standard Feynman diagram
methods for the calculation of the basic quantities
in perturbation theory such as one-loop scattering amplitudes.
The renewed interest in this approach
is largely due to the work of Bern and Kosower,
who derived new rules for the construction
of one-loop QED/QCD 
and supergravity amplitudes by analyzing the infinite string tension
limits of the corresponding amplitudes in appropriate
string models ~\cite{BernKosower}. 
Those ``Bern-Kosower Rules'' were applied successfully 
to the calculation of five-gluon amplitudes
\cite{bediko} as well as four-graviton amplitudes
\cite{fourgrav}.
Strassler \cite{strassler} succeeded in
rederiving those same ``Bern-Kosower Rules'',
without the use of string theory,
by representing the QED/QCD one-loop effective action
in terms of suitable first-quantized
path integrals, and evaluating them
in a way analogous to string theory ~\cite{polbook}
\footnote{In the nonabelian case 
the arguments of \cite{strassler} fall somewhat
short of a complete rederivation of the
Bern-Kosower rules.}.
This approach to the Bern-Kosower formalism
has been further developed by various authors
\cite{ss2,ss3,dashsu,sato2,shaisultanov,rolsat,sato}. 
It turned out to be very effective for
the calculation of photon amplitudes in QED, in
particular for problems involving constant external
fields \cite{rescsc,adlsch,cadhdu,gussho,sandansky}.

A similar formalism was developed in an independent line
of work starting from the heat kernel representation
of quantum field theory propagators \cite{mckeon,mckreb,dilmck}.
For a discussion
of the differences between both approaches see \cite{MnW}.

Until very recently particle path integral representations
for one-loop effective actions were known essentially only
for scalar and for (vector-) gauge theories. Generalizations
to Yukawa and axial couplings were constructed heuristically
in \cite{Mont}, and more rigorously and completely
in \cite{DHokerGagne}, where the general case of a
spin $\half$ - loop coupled to background scalar,
pseudoscalar, vector, axial-vector, and antisymmetric
tensor fields was treated. However in both cases
the construction of the path integrals had to be done
separately for the real and the imaginary parts of the
(Euclidean) effective action. Moreover, for the imaginary part
certain ``insertion operators'' had to be defined in addition
to the worldline Lagrangian. 
In part I of the present series \cite{axialI} 
the case of a vector - axialvector background was considered,
and a path integral representation for the corresponding
one-loop effective action derived which avoids both of
these complications. Let us repeat this derivation shortly.

One-loop axialvector and vector current correlators can be extracted 
from the determinant of the 
non-Hermitian operator $H-im$ where 
\begin{equation}
\label{e1}
H = 
p \!\! / + A \!\!\! / + \gamma_5 {A \!\!\! / }_5
\, , 
\hspace{1cm} (p = -i \partial)
\end{equation}
(in Euclidean space)~\cite{Gaume,r62}.  
It has proved possible to deal with 
this functional determinant without resorting to a decomposition
into its real and imaginary parts~\cite{Gaume,r62}.
The basic identity
used in our present approach is
\cite{Italian}

\begin{eqnarray}
\label{e2}
\left( 
p \!\! / + A \!\!\! / + \gamma_5 {A \!\!\! / }_5
\right)^2
& = & \left( p_\mu + A_\mu - \gamma_5 \sigma_{\mu\nu} A^\nu_5 \right)^2 
+ (D-2) A_5^2 \nonumber \\
& & + i A^\nu_{5,\nu} \gamma_5 
- \frac{i}{2} \sigma_{\mu\nu} 
\left( \partial^\mu A^\nu - \partial^\nu A^\mu \right) \, 
\end{eqnarray}\no
valid for any even dimension
$D$ in Euclidean spacetime%
\footnote{%
We work in the Euclidean throughout with
a positive definite metric
$g_{\mu\nu}={\,\mathrm diag}(++\ldots +)$.
Our Dirac matrix conventions are
(with some abuse of notation)
$\lbrace\gamma_{\mu},\gamma_{\nu}\rbrace = 2g_{\mu\nu}$,
$\gamma_5^2 = \Eins$, 
$\gamma_{\mu,5}^{\dag} = \gamma_{\mu,5}$,
$\sigma_{\mu\nu} = 
\frac{1}{2} \left[ \gamma_\mu , \gamma_\nu \right]$,
$\mu,\nu = 1,\ldots,D$.
The Euclidean $\varepsilon$ - tensor is defined by
$\varepsilon_{1\cdots D} = +1$.  We do not differentiate
between superscripts and subscripts.
The corresponding Minkowski space formulas
are obtained by $g_{\mu\nu}\rightarrow \eta_{\mu\nu}
= {\,\mathrm diag}(-+\ldots +)$,$
k_D\rightarrow -ik^0, T\rightarrow is,
\varepsilon^{\mu_1\cdots \mu_D}
\rightarrow
i\varepsilon^{\mu_1\cdots \mu_D},
\varepsilon^{0\cdots (D-1)} = +1
$.
}.
Our object of interest is the heat kernel
\begin{equation}
\label{e3}
K(y,z;T) 
= \langle y | e^{- H^2 T } | z \rangle \,
\, .
\end{equation}\no
Standard arguments~\cite{FeynmanHibbs} 
show that we can represent $K$ as a path-ordered
quantum mechanical path integral as follows,

\begin{equation}
\label{e4}
K(y,z;T) = \int^{x(T) = y}_{x(0)= z} \D x(\tau)
\P \exp \left[ - \int^T_0 \L \left(x,\gamma\right) d\tau \right]
\end{equation}
where
\begin{eqnarray}
\label{L}
\L(x,\gamma) & = & 
\frac{ {\dot x}^2 }{4} + i A \cdot \dot x
- \frac{i}{4} \left[ \gamma^\mu, \gamma^\nu \right]
\left( \partial^\mu A^\nu - \partial^\nu A^\mu \right) \nonumber \\
& & - \frac{i}{2} \gamma_5 \left[ \gamma^\mu, \gamma^\nu \right] 
A^\nu_5 {\dot x}^\mu 
+ i \gamma_5 \partial_\mu A^\mu_5 + (D-2) A_5^2 \, .
\end{eqnarray}
In ~\cite{axialI} only the effective action was considered,
which is obtained from the heat-kernel $K$ by
taking the functional trace and integrating over
proper-time. The coherent state method \cite{DHokerGagne,Kashiwa} was
then used to transform the trace over spinor indices
into a Grassmann path integral, with the result

\bear
\Gamma[A,A_5]&=&
-\half
\int_0^{\infty}{dT\over T}
\,\e^{-m^2T}
\Dx
\int
{\cal D}\psi
\,\,\e^{-\int_0^Td\tau\, L(\tau)}\label{GammaAA5}\\
L &=& 
\kinb + \half\psi\cdot\dot\psi 
+i\dot x\cdot A
-i \psi\cdot F\cdot\psi
\nonumber\\&&
 +i\hat\gamma_5
\Bigl[
-2\dot x\cdot\psi\psi\cdot A_5
+\partial\cdot A_5
\Bigr]
+ (D-2) A_5^2
\label{defL}
\ear\no
Although our use of the coherent state method completely parallels 
\cite{DHokerGagne} we
spell out the details of this transformation 
in appendix \ref{coherentstates} 
for completeness
(for $D=4$).
The result is a formal correspondence
$[\gamma^{\mu},\gamma^{\nu}] \rightarrow 4 
\psi^{\mu}\psi^{\nu}$, 
$\gamma_5 \rightarrow \hat \gamma_5$.
This formula for the effective action is valid for any
even spacetime dimension, and also for the massive case.
For fixed $T$ one has to calculate a
coordinate path integral $\int {\cal D}x$ over the space of all closed
loops in 
spacetime with periodicity $T$, and a Grassmann
path integral $\int {\cal D}\psi$.
The boundary conditions on the Grassmann path integral are,
after expansion of the interaction exponential, determined by the
power of $\hat\gamma_5$ appearing in a given term; they are
(anti) periodic with period $T$ if that power is even (odd).
After the boundary conditions are determined $\hat\gamma_5$ can be
replaced by unity.

In part I it was shown how to apply this path integral to
a calculation of the effective action itself in
the heat kernel expansion, both for its anomalous and
non-anomalous parts. 
In the present paper we exploit
it for deriving master formulas for the one-loop (off-shell)
amplitudes with an arbitrary number of vector and axial-vector legs.
Those are extensions of the string-inspired master formula
for QED photon scattering, which in turn is a special case
of the QCD Bern-Kosower master formula \cite{BernKosower}.

We then investigate alternative methods for computing
the general element of the heat kernel.  Although these methods
apply to both diagonal and off-diagonal calculations,
for the latter we must restrict our attention to the massless fermion
since a mass parameter turns out to be problematic.
We use the Fradkin-Gitman (FG) formalism~\cite{FG} 
to eliminate the Dirac matrices in (\ref{e4}) in favour of
Grassmann parameters; 
for this we recall the approach of 
\cite{TonyMe}.  
We then demonstrate how both
$
\P \exp \int^T_0 d\tau \, A^{\mu\nu}(\tau) \sigma^{\mu\nu}
$
and
$
\P \exp \int^T_0 d\tau \, \vec{A}(\tau) \cdot \vec{\sigma}
$
can be written as closed linear combination of spin matrices,
thereby completely circumventing the need for introducing
any Grassmann integration.  
The latter path-ordered operator is relevant
in connection with our observation that the
matrix valued Lagrangian appearing in (\ref{e4}) commutes with
the chirality operator $\gamma_5$.  We exploit this property
by showing how 
the decomposition of the $SO(4)$ generators
$\sigma_{\mu\nu}$ appearing in (\ref{e4}) into the generators of the 
$SU(2)$ subgroups by use of the 't Hooft 
symbols $\eta^{\pm}_{a\mu\nu}$ (originally introduced in the context
of instantons~\cite{abc}) can simplify the functional
integral considered in (\ref{e4}).  
The utility of this approach is illustrated by a calculation of the 
axial anomaly.

We also demonstrate that when the Dirac structure of $K$ is decomposed
using the FG approach, the resulting
zero-mode dependence leads to a very natural
functional interpretation of 
the ``Feynman rules'' occurring in the coherent state method.
We show how
external fermions can be integrated into the FG approach
and how this leads to a suitable unification
of the background axial, vector and spinor cases.
The possibility of handling isospin in the same way 
that we have treated ordinary spin
is also considered briefly.
For the pure vector case we also examine
the interplay between worldline supersymmetry
and the fermionic boundary conditions,
using a theorem from
$0+1$ dimensional topological field theory.

\section{A Master formula for the one-loop vector - axialvector amplitudes}
\label{SecMaster}

Generally,
as a first step in any evaluation of the double path integral
in (\ref{GammaAA5})
one has to eliminate the zero mode(s)
contained in it. For $\int{\cal D}x$ this is done by fixing the
average position of the loop, i.e. one writes

\bear
x^{\mu}(\tau) &=& x_0^{\mu} + y^{\mu}(\tau)\nonumber\\
\Dx &=& \int dx_0 \Dy
\label{split}
\ear\no
where

\begin{equation}
x_0^{\mu}\equiv {1\over T}\int_0^T d\tau\, x^{\mu}(\tau)
\label{defx0}
\end{equation}
\no
The Grassmann path integral has a zero-mode
only in the periodic case.
This must again be separated out,

\bear
\psi^{\mu}(\tau) &=& \psi_0^{\mu} + \xi^{\mu}(\tau)
\label{splitgrass}\\
\int_0^Td\tau \, \xi^{\mu}(\tau) &=& 0
\label{xicond}
\ear\no
The zero mode integration then produces the
$\varepsilon$ - tensor expected for a spinor loop
with an odd number of axial insertions,

\be
\int d^D\psi_0
\psi_0^{\mu_1}\psi^{\mu_2}_0\cdots\psi^{\mu_D}_0
=\varepsilon^{\mu_1\cdots\mu_D}
\label{zeromodeintegral}
\ee\no
The remaining $y$ - and $\xi$ -
path integrals have invertible kinetic terms.
In the ``string-inspired'' formalism they are
performed using worldline Green's functions adapted to the
boundary conditions,

\begin{eqnarray}
\langle y^\mu(\tau_1)\,y^\nu(\tau_2)\rangle
&=& -g^{\mu\nu}{G}_{B}(\tau_1,\tau_2)\nonumber\\
\langle\psi^{\mu}(\tau_1)\, \psi^{\nu}(\tau_2)\rangle_A
&=& 
g^{\mu\nu}\half {G}_{F}(\tau_1,\tau_2)\nonumber\\
\langle\xi^{\mu}(\tau_1)\, \xi^{\nu}(\tau_2)\rangle_P
&=& 
g^{\mu\nu}\half {\dot{G}}_{B}(\tau_1,\tau_2)
\label{modcorrelators}
\end{eqnarray}\no
where 
\footnote{There is a certain freedom in the choice
of the bosonic correlator. Our $G_B$ is the
one generally used in the ``string-inspired''
formalism. It yields the same final results as the 
one employed in \cite{TonyMe}, as discussed in \cite{MnW}.}

\begin{eqnarray}
G_B(\tau_1,\tau_2)
    &=& \mid \tau_1 - \tau_2\mid -
{{(\tau_1 - \tau_2)}^2\over T}\nonumber\\
\dot G_B(\tau_1,\tau_2)
&=&
{\rm sign}(\tau_1-\tau_2)
-2 {\tau_1-\tau_2 \over T}\nonumber\\
G_F(\tau_1,\tau_2)
    &=& {\rm sign} (\tau_1-\tau_2)
\label{defGBGF}
\end{eqnarray}
\no
A ``dot'' always refers to a derivative in the
first variable. 
The last information needed are the free Gaussian path integral
determinants, which with the present conventions are
\footnote{When comparing to \cite{axialI}
note that we have shifted a factor of $2^{D\over 2}$ from the overall
factor to the fermionic path integral.}

\bear
\int{\cal D}y
\,\e^{-\int_0^T d\tau\, \fourth
\dot y^2}
&=& {(4\pi T)}^{-{D\over 2}} \nonumber\\
\int_A{\cal D}\psi\, \e^{-\int_0^Td\tau \,\half\psi\cdot\dot\psi}
&=& 2^{D\over 2} =: N_A\nonumber\\
\int_P{\cal D}\xi\, \e^{-\int_0^Td\tau \,\half\xi\cdot\dot\xi}
&=& (-1)^{D\over 2} =: N_P
\label{freepi}
\ear\no

To extract the scattering amplitude from the effective action, as usual
we must specialize the background fields to plane waves,
and then keep the part of the effective action which is linear
in all polarization vectors. 
As a preliminary step, it is convenient to linearize the
term quadratic in $A_5$ by introducing an auxiliary path
integration, writing

\bear
\label{linearizeA52}
\exp \Bigl[-(D-2)\int_0^Td\tau A_5^2\Bigr]
&=&
\int {\cal D}z \exp \Bigl[-\int_0^Td\tau 
\Bigl({z^2\over 4}+i\sqrt{D-2}z\cdot A_5\Bigr)
\Bigr]
\ear\no
This allows us to define an axial-vector vertex operator
as follows,

\bear
V_{A_5}[k,\varepsilon] &\equiv&
\int_0^Td\tau
\Bigl(i\varepsilon\cdot k 
+ 2\varepsilon\cdot\psi
\dot x\cdot\psi
+ \sqrt{D-2}\,\varepsilon\cdot z
\Bigr)
\,\e^{ik\cdot x}
\label{defaxvectvertop}
\ear\no
Before using this vertex operator for Wick-contractions, as usual
it is convenient to formally rewrite it as a
linearized exponential. This can be done introducing 
Grassmann variables,

\bear
V_{A_5}[k,\varepsilon] 
&=&
\int_0^Td\tau
\int d\theta
\exp\Bigl\lbrack ik\cdot x + i\theta \varepsilon\cdot k
+\sqrt{2}\varepsilon\cdot\psi
+\sqrt{2}\theta\dot x\cdot\psi
\non\\&&\quad\quad
+\sqrt{D-2}\,\theta\varepsilon\cdot z
\Bigr\rbrack
\Large\mid_{{\rm lin}(\varepsilon)}
\label{rewriteaxvectvertop}
\ear\no
Here $\theta$ is a Grassmann variable with $\int d\theta\theta =1$,
and $\varepsilon$ must now also be formally treated as Grassmannian.
The vectors are represented by the usual photon vertex
operator familiar from string theory

\bear
V_A[k,\varepsilon]&\equiv&
\int_0^Td\tau
\Bigl[
\varepsilon\cdot \dot x
+2i
\varepsilon\cdot\psi
k\cdot\psi
\Bigr]
\,\e^{ik\cdot x}
\non\\
&=&
\int_0^Td\tau\int d\theta
\exp\Bigl\lbrack
ik\cdot(x+\sqrt{2}\theta\psi)
+\varepsilon\cdot (-\theta\dot x +\sqrt{2}\psi)
\Bigr\rbrack
\Large\mid_{{\rm lin}(\varepsilon)}
\label{rewritephotonvertop}
\ear
\noindent\no
which we have also rewritten as a linearized exponential.
Those definitions allow us to represent the
one-loop amplitude with $M$ vectors and $N$ axialvectors
in the following way,

\bear
\Gamma[\lbrace k_i,\varepsilon_i\rbrace,
\lbrace k_{5j},\varepsilon_{5j}\rbrace]
&=&
-\half N_{A,P}(-i)^{M+N}
\Tintm 
\PITD
\non\\
&&\Mneg
\Mneg
\times
\Bigl\langle
V_{A}[k_1,\varepsilon_1]\ldots
V_{A}[k_M,\varepsilon_M]
V_{A_5}[k_{51},\varepsilon_{51}]\ldots
V_{A_5}[k_{5N},\varepsilon_{5N}]
\Bigr\rangle
\label{repMvectorNaxial}
\ear\no
where the global sign refers to the ordering
$\varepsilon_1\varepsilon_2\ldots\varepsilon_M
\varepsilon_{51}\varepsilon_{52}\ldots\varepsilon_{5N}$
of the polarisation vectors.
It is then straightforward to perform the bosonic path integrations,

\bear
&&
\int {\cal D}x\int{\cal D}z\,
V_{A}[k_1,\varepsilon_1]\ldots
V_{A_5}[k_{5N},\varepsilon_{5N}]
\,\e^{-\int^T_0 d\tau \Bigl({{\dot x}^2\over 4} + {z^2\over 4}\Bigr)}
=
\non\\&&
\PITD 
\int_0^T d\tau_1 \cdots \int d\theta_M
\int_0^T d\tau_{51} \cdots \int d\theta_{5N}
\non\\&&\times
\exp\biggl\lbrace
\half
G_{BIJ}K_I\cdot K_J
-i\theta_i\dot G_{BiJ}\varepsilon_i\cdot K_J
-i\sqrt{2}\theta_{5i}\dot G_{BiJ}\psi_i\cdot K_J
\non\\&&
-\half \ddot G_{Bij}\theta_i\theta_j\varepsilon_i\cdot\varepsilon_j
-\sqrt{2}\ddot G_{Bij}\theta_i\theta_{5j}\varepsilon_i\cdot\psi_j
-\ddot G_{Bij}\theta_{5i}\theta_{5j}\psi_i\cdot\psi_j
\non\\&&
+\sqrt{2}(\varepsilon_i\cdot\psi_i +\varepsilon_{5j}\cdot\psi_j)
+\sqrt{2}i\theta_i k_i\cdot\psi_i
+ i\theta_{5i}\varepsilon_{5i}\cdot k_{5i} 
\non\\&&
-(D-2)\delta(\tau_i-\tau_j)\theta_{5i}\theta_{5j}
\varepsilon_{5i}\cdot\varepsilon_{5j}
\biggr\rbrace
\Large\mid_{{\rm lin}(\lbrace\varepsilon_i\rbrace;
\lbrace\varepsilon_{5j}\rbrace)}
\label{resdxdz}
\ear\no
Here and in the following all small repeated indices run 
over either the vector or the axial vector indices,
while capital repeated indices run over all of them
($\lbrace K_I\rbrace $ denotes the set of all external momenta).
The remaining
$\psi$ - path integral is still Gaussian. For its performance
we must
now distinguish between even and odd numbers of
axialvectors. For the antiperiodic case, $N$ even,
there is no zero-mode, and the integration 
can still be done in closed form.
The only complication is the
existence of the term 
$-\ddot G_{Bij}\theta_{5i}\theta_{5j}\psi_i\cdot\psi_j$.
It modifies the worldline propagator $G_F$ to

\bear
G_{F12}^{(N)} 
&\equiv&
2\langle \tau_1\mid
\Bigl(\partial + 2 B^{(N)}\Bigl)^{-1}
\mid\tau_2\rangle
\label{defGN}
\ear\no
where $B^{(N)}$ denotes the operator with integral kernel

\bear
B^{(N)}(\tau_1,\tau_2) &=&
\delta(\tau_1 - \tau_i)\theta_{5i}
\ddot G_{Bij}
\theta_{5j}\delta(\tau_j-\tau_2)
\label{defBN}
\ear\no
($B^{(N)}$ acts trivially on the Lorentz indices, which we
suppress in the following).
Expanding the right hand side of eq.(\ref{defGN}) 
in a geometric series and resumming one obtains a
matrix representation for $G_{F12}^{(N)}$, 

\bear
G_{F}^{(N)} 
&=&
{G_{F}
\over
\Eins +\ddot\Theta G_F}
=
G_F -G_F\ddot\Theta G_F + \ldots
\label{GNexplicit}
\ear\no
Here $\ddot\Theta_{ij}$ is the antisymmetric
$N\times N$ matrix with entries
$\theta_{5i}\ddot G_{Bij}\theta_{5j}$ (no summation).
Moreover, the fermionic path integral determinant
changes by a factor 

\bear
\Det^{\half}\Bigl(\Eins + 2B^{(N)}\partial^{-1}\Bigr)
&=&
{\det}
\bigl(\Eins +\ddot\Theta G_F\bigr)^{{D\over 2}} 
\label{detfact}
\ear\no
as is easily seen using the $ln det = tr ln$ - formula
(note that on the left hand side we have a functional determinant,
on the right hand side the determinant of a $N\times N$ matrix).
Using these results the fermionic path integral can be
eliminated, yielding the following master formula for
this amplitude,

\bear
\Gamma_{\rm even}[\lbrace k_i,\varepsilon_i\rbrace;
\lbrace k_{5j},\varepsilon_{5j}\rbrace]
&=&
-{N_A\over 2}(-i)^{M+N}
\Tintm \PITD
\non\\
&&\Mneg\Mneg\Mneg\times
\int_0^Td\tau_1\int d\theta_1
\cdots
\int_0^Td\tau_{5N}\int d\theta_{5N}
\,{\det}
\bigl(\Eins +\ddot\Theta G_F\bigr)^{{D\over 2}} 
\non\\
&&\Mneg\Mneg\Mneg\times
\exp\biggl\lbrace
\half
G_{BIJ}K_I\cdot K_J
-i\theta_i\dot G_{BiJ}\varepsilon_i\cdot K_J
-\half \ddot G_{Bij}\theta_i\theta_j\varepsilon_i\cdot\varepsilon_j
\non\\
&&\Mneg\Mneg
- {G_{Fij}^{(N)}\over 2}
\Bigl(\varepsilon_i+i\theta_i k_i
+\varepsilon_{5i}
-i\theta_{5i}\dot G_{BiR}K_R
+\ddot G_{Bir}\theta_{5i}\theta_r\varepsilon_r
\Bigr)
\non\\ && \Mneg\Mneg
\cdot
\Bigl(\varepsilon_j+i\theta_j k_j
+\varepsilon_{5j}
-i\theta_{5j}\dot G_{BjS}K_S
+\ddot G_{Bjs}\theta_{5j}\theta_s\varepsilon_s
\Bigr)
+i\theta_{5i}\varepsilon_{5i}\cdot k_{5i}
\non\\ && \Mneg\Mneg
-(D-2)\delta(\tau_i-\tau_j)\theta_{5i}\theta_{5j}
\varepsilon_{5i}\cdot\varepsilon_{5j}
\biggr\rbrace
\Large\mid_{{\rm lin}(\lbrace\varepsilon_i\rbrace;
\lbrace\varepsilon_{5j}\rbrace)}
\label{axialevenmaster}
\ear
As is usual in the string-inspired formalism all momenta are
{\sl outgoing}.
Let us verify the correctness of this formula for the 
case of the massive 2-point axialvector function
in four dimensions. 
Expanding out the exponential as well as the 
determinant factor, and performing the 
two 
$\theta$ - integrals,
we obtain the following parameter integral,

\bear
\Gamma[k_1,\varepsilon_1;k_2,\varepsilon_2]
&=&
2\Tintm \PITD 
\int_0^Td\tau_1\int_0^T d\tau_2
\non\\&&\Mneg\times
\e^{G_{B12}k_1\cdot k_2}
\biggl\lbrace
2(D-2)\delta(\tau_1-\tau_2)\varepsilon_1\cdot\varepsilon_2
-(D-1)\ddot G_{B12}G_{F12}^2\varepsilon_1\cdot\varepsilon_2
\non\\&&\Mneg
-G_{F12}^2\dot G_{B12}^2
(\varepsilon_1\cdot\varepsilon_2 k_1\cdot k_2 -
\varepsilon_1\cdot k_1 \varepsilon_2\cdot k_2)
-\varepsilon_1\cdot k_1 \varepsilon_2\cdot k_2
\biggr\rbrace
\label{A5A5int}
\ear\no
As usual in this type of calculation we rescale 
$\tau_{1,2}=T u_{1,2}$ , and use the
translation invariance in $\tau$ to
set $u_2=0$. 
Setting also $k=k_1=-k_2$ this leads to

\bear
\Gamma^{\mu\nu}[k]
&=&
2\Tintm \PITD 
\biggl\lbrace
2(D-2)Tg^{\mu\nu}
-2(D-1)Tg^{\mu\nu}
\non\\&&\hspace{-50pt}
+ \int_0^1 du
\,\e^{-Tu(1-u)k^2}
\Bigl[
2(D-1)Tg^{\mu\nu}
+(1-2u)^2T^2(g^{\mu\nu}k^2-k^{\mu}k^{\nu})
+T^2k^{\mu}k^{\nu}
\Bigr]
\biggr\rbrace
\label{A5A5int2}
\ear\no
In the massless case
the first two terms in braces
do not contribute in dimensional regularization,
since they are of tadpole type. For the remaining
terms both integrations are elementary,
and the result is, using
$\Gamma$ - function identities,
easily identified with the
standard result for the
massless QED vacuum polarisation.
A suitable integration by part verifies
the agreement with field theory also for
the massive case.
Here the tadpole terms do contribute,
and the comparison
shows that to get the precise $D$ - dependence of
the amplitude, appropriate to dimensional
regularisation using an anticommuting $\gamma_5$,
it was essential to keep the explicit $D$ - dependence
of the $A_5^2$ - term in the worldline Lagrangian.

For an odd number of axial vectors, we need to go back to
eq.(\ref{resdxdz}) and replace $\psi$ by $\psi_0 +\xi$.
The ${\cal D}\xi$ - path integral is then executed in the
same way as before, but with the propagator $G_F$ changed
to $\dot G_B$. The final result becomes

\bear
\Gamma_{\rm odd}[\lbrace k_i,\varepsilon_i\rbrace;
\lbrace k_{5j},\varepsilon_{5j}\rbrace]
&=&
-{N_P\over 2}(-i)^{M+N}
\Tintm \PITD
\non\\
&&\Mneg\Mneg\times
\int_0^Td\tau_1\int d\theta_1
\cdots
\int_0^Td\tau_{5N}\int d\theta_{5N}
\,{\det}
\bigl(\Eins +\ddot\Theta \dot G_B\bigr)^{{D\over 2}} 
\int d^D\psi_0
\non\\
&&\Mneg\Mneg\times
\exp\biggl\lbrace
\half
G_{BIJ}K_I\cdot K_J
-i\theta_i\dot G_{BiJ}\varepsilon_i\cdot K_J
-\half \ddot G_{Bij}\theta_i\theta_j\varepsilon_i\cdot\varepsilon_j
\non\\
&&\Mneg\Mneg
- {\dot G_{Bij}^{(N)}\over 2}
\Bigl(\varepsilon_i+i\theta_i k_i
+\varepsilon_{5i}
-i\theta_{5i}\dot G_{BiR}K_R
+\ddot G_{Bir}\theta_{5i}\theta_r\varepsilon_r
+\sqrt{2}\ddot G_{Bir}\theta_{5i}\theta_{5r}\psi_0
\Bigr)
\non\\ && \Mneg\Mneg
\cdot
\Bigl(\varepsilon_j+i\theta_j k_j
+\varepsilon_{5j}
-i\theta_{5j}\dot G_{BjS}K_S
+\ddot G_{Bjs}\theta_{5j}\theta_s\varepsilon_s
+\sqrt{2}\ddot G_{Bjs}\theta_{5j}\theta_{5s}\psi_0
\Bigr)
\non\\ && \Mneg\Mneg
+i\theta_{5i}\varepsilon_{5i}\cdot k_{5i}
-(D-2)\delta(\tau_i-\tau_j)\theta_{5i}\theta_{5j}
\varepsilon_{5i}\cdot\varepsilon_{5j}
-i\sqrt{2}\theta_{5i}\dot G_{BiJ}\psi_0\cdot K_J
\non\\ && \Mneg\Mneg
-\sqrt{2}\ddot G_{Bij}\theta_i\theta_{5j}\varepsilon_i\cdot\psi_0
+\sqrt{2}\bigl(\sum\varepsilon_i +\sum\varepsilon_{5j}\bigr)\cdot\psi_0
+\sqrt{2}i\theta_i k_i\cdot\psi_0
\biggr\rbrace
\Large\mid_{{\rm lin}(\lbrace\varepsilon_i\rbrace;
\lbrace\varepsilon_{5j}\rbrace)}
\non\\
\label{axialoddmaster}
\ear\no
Here $\dot G_B^{(N)}$ is defined analogously to eq.(\ref{GNexplicit}),

\bear
\dot G_{B}^{(N)} 
&=&
{\dot G_{B}
\over
\Eins +\ddot\Theta \dot G_B}
\label{GdotNexplicit}
\ear\no
The integrand still depends on the zero-mode $\psi_0$, which
is to be integrated according to eq.(\ref{zeromodeintegral}).
For a given dimension of space-time this can, of course, be
easily worked out explicitly.

\section{The Dirac algebra using FG and deducing spin factors}

Of course it is desirable to have a 
representation for the heat-kernel 
which applies to both diagonal and off-diagonal elements.
The presence of a non-vanishing fermion mass in the latter case
becomes an essential complication which we will not consider 
in the present paper.
Moreover we specialize to $D=4$,
since we are now going to use more specific properties of
the Dirac algebra, 
although many of our methods have analogues in other dimensions.
For this generalization the
coherent state method seems less convenient, and
we rather use the ``Weyl symbol'' formalism
\cite{bermar,FG,TonyMe}, as follows.
We begin by presenting two equations of which we will
make frequent use \cite{FG,Gitman98,Fried,Vasiliev}:
\begin{equation}
\label{e5}
\P \exp \int^T_0 d\tau f \left( \tau , \gamma \right)
= \Phat \left. 
\exp \int^T_0 d\tau f\left( \tau , \case{\partial}{\partial \rho}
\right) \; \P \exp \int^T_0 d\tau \rho(\tau) \cdot \gamma \right|_{\rho=0}
\end{equation}
\begin{eqnarray}
\label{e6}
\P \exp \int^T_0 d\tau \rho(\tau) \cdot \gamma
& = & \exp \left[ - \frac{1}{2} \int^T_0 \sign(\tau_1 - \tau_2)
\rho(\tau_1) \cdot \rho(\tau_2) \, d\tau_1 d\tau_2 \right] \\
& & \times \exp \int^T_0 d\tau \rho(\tau) \cdot \gamma \nonumber
\end{eqnarray}
where $\rho^\mu(\tau)$, with $\mu=1, \ldots ,4$, are Grassmann variables
which anticommute%
\footnote{The anticommutation of $\gamma$ with $\rho$ (and $\theta$ 
to be introduced later) is a matter
of algebraic convenience.  It should be possible to formulate
the problem with the 
assumption that $\gamma$ is a c-number valued matrix,
in which case $\Phat$ would have to denote {\em reversed} path-ordering.  
However,
we follow \cite{FG} for convenience and retain the convention that
$\gamma$ is a Grassmann quantity.} 
with $\gamma^\nu$,
and $f(\tau,\gamma)$ is a linear combination
of antisymmetrized products of $\gamma$-matrices.
Path
ordering of matrices (both $\gamma$ and later $\vec{\sigma}$),
is denoted by $\P$ while $\Phat$ denotes path 
ordering of Grassmann derivatives $\frac{\partial}{\partial \rho}$.

Equations (\ref{defGBGF}), (\ref{e5}) and (\ref{e6}) 
can be used to convert
(\ref{e4}) to
\begin{eqnarray}
\label{e7}
K(y,z;T) & = & \int^{x(T)=y}_{x(0)=z} \D x(\tau) 
\Phat \exp 
\int^T_0 d\tau \L \left( x, \case{\partial}{\partial \rho} 
\right)
\nonumber \\
& & \left. \times \exp \left[ - \frac{1}{2} \int^T_0 d\tau_1 d\tau_2 
\, G_F(\tau_1,\tau_2) 
\rho(\tau_1) \cdot \rho(\tau_2) \right] \exp \int^T_0 d\tau \rho(\tau) \cdot 
\gamma 
\right|_{\rho = 0}
\end{eqnarray}
In this equation,
the $\Phat$ operator above is actually superfluous since the Lagrangian
$\L$ is even in the anticommuting variables. 

Notice that 
the last exponential in equations (\ref{e6}) and (\ref{e7}) depends only on
the finite set of so-called
{\em zero modes}
of the Grassmann field, 
$T \zero \rho^\mu = \int^T_0 d\tau \rho^\mu(\tau)$, so the 
Taylor series truncates.

Equation~(\ref{e6}) is a special case of Wick's theorem
which relates the $\P$ and {\em Sym} orderings of the operator
$\exp \int^T_0 d\tau \rho \cdot \gamma$.  
A more extensive discussion can be found in
\cite{Gitman98}.
The relation follows from
generalized
form of Wick's theorem (see, for instance, page 60 of \cite{Polchinski})
which can be used to relate any two operator orderings of 
a given functional. 
It's 
appearance in (\ref{e7}) allows us to draw strong connections with the 
Feynman rules used in the string-inspired formalism, 
enumerated in chapter~\ref{SecMaster}. 
For this we defer the reader to appendix \ref{NoPI}.

One method of dealing with (\ref{e4}) is to 
use the Fradkin-Gitman approach \cite{FG} to replace the time ordered
exponential involving Dirac matrices by an integral over Grassmann variables.
The result is
\begin{equation}
\label{e15}
K(y,z;T)  
\propto 
\left. \exp \left( i \gamma \cdot \frac{\partial}{\partial \theta} \right)
\int^{x(T)=y}_{x(0)=z} \D x(\tau) \int_{\psi(0) + \psi(T) = \sqrt{2} \theta}
\D\psi(\tau) 
e^{\frac{1}{2} \psi(T) \cdot \psi(0) + 
\int^T_0 d\tau L(\tau)}
\right|_{\theta = 0} \nonumber \, ,
\end{equation}
where $L(\tau)$ is defined in (\ref{defL}).
This approach has been exploited in~\cite{TonyMe}.

However, it is a bit awkward to work with (\ref{e15}), due to
having the quantity $\theta$ arising in the boundary condition
on the Grassmann parameter $\psi$ over which we are integrating.
We can simplify (\ref{e4}) directly by essentially generalizing a
well-known functional identity \cite{Fried}.  We start by noting the
standard result
\begin{equation}
\label{e16}
\int \D \omega_i \exp\left[ \frac{1}{2} \omega_i A_{ij} \omega_j
+ \rho_i \omega_i \right ] 
= {\det}^{1/2} A \exp \frac{1}{2} \rho_i A^{-1}_{ij} \rho_j \, .
\end{equation}
where $\omega_i$ and $\rho_i$ are Grassmann and $A_{ij} = - A_{ji}$.
{}From (\ref{e16}), it follows that
\begin{eqnarray}
Q & \equiv & 
\left. \exp \left[ \frac{1}{2} \frac{\partial}{\partial \rho_i} 
B_{ij} \frac{\partial}{\partial \rho_j} \right] 
\exp \left[ \frac{1}{2} \rho_i A^{-1}_{ij} \rho_j + \rho_i c_i \right]
\right|_{\rho=0} \nonumber \\
&& = \exp \left[ \frac{1}{2} \frac{\partial}{\partial \rho_i}
B_{ij} \frac{\partial}{\partial \rho_j} \right] {\det}^{-1/2} A \\
& & \hspace{1cm}
\left. \int \D \omega_i \exp \left[ \frac{1}{2} \omega_i A_{ij} \omega_j
+ \rho_i \left( \omega_i + c_i \right) \right] \right|_{\rho=0}
\nonumber 
\end{eqnarray}
(where $c_i$ is Grassmann and $B_{ij} = - B_{ji}$),  
so that  
\begin{equation}
\label{e18}
Q = {\det}^{1/2} A^{-1} \int \D \omega_i \exp \left[ \frac{1}{2}
\omega_i \left( A_{ij} + B_{ij} \right) \omega_j + c_i B_{ij} \omega_j
+ \frac{1}{2} c_i B_{ij} c_j \right] \, .
\end{equation}
Again, employing (\ref{e16}) to compute the functional integral in (\ref{e18}),
we obtain,
\begin{eqnarray}
Q & = & {\det}^{1/2} \left( \Eins + A^{-1} B \right)
\exp \left[ - \frac{1}{2} c_i B_{ij} \left( 1+ A^{-1} B \right)^{-1}_{jk}
\left( A^{-1} B \right)_{kl}
c_l + \frac{1}{2} c_i B_{ij} c_j \right] \label{e19a} \label{e19} \\
& = & {\det}^{1/2} \left( \Eins + A^{-1} B \right)
\exp \left[ \frac{1}{2} c_i \left(B (A+B)^{-1} B \right)_{ij} c_j \right]
\label{e19b}
\end{eqnarray}
Although equation~(\ref{e19b}) is formally simpler,
equation~(\ref{e19a})
turns out to be more convenient
for our (perturbative) calculations below.
Remembering that $c_i$ in (\ref{e19}) is a Grassmann variable, we
see that the second exponential
in (\ref{e19}) can be expanded in a finite series provide there
are a finite number of $c_i$'s.  

The result~(\ref{e19a}) is useful for treating the 
dependence of the path integral in equation~(\ref{e4}) on
the gamma matrices.  To see this, 
we now 
consider the functional
\begin{equation}
I_4[B_4] = \P \exp \frac{1}{2} \int^T_0 d\tau B^{\mu\nu}_4(\tau) \gamma^{\mu}
\gamma^{\nu} 
\end{equation}
which is relevant
for the case at hand (\ref{e4}) if
the antisymmetric tensor $B_4(\tau)$
is given by
$$
B^{\mu\nu}_4 = -\case{i}{2} F^{\mu\nu} + \case{i}{2}
\epsilon^{\mu\nu\lambda\sigma} \dot x ^\lambda A^\sigma_5 \, .
$$  
We write,
using (\ref{e5})
and (\ref{e6}),
\begin{eqnarray}
I_4[B_4] 
 & = & \exp \int^T_0 d\tau \frac{1}{2} \left[ 
B^{\mu\nu}_4(\tau) \frac{\partial}{\partial \rho^{\mu}}
\frac{\partial}{\partial \rho^\nu} \right] \\
&& \times \exp \left[ - \frac{1}{2} \int^T_0 d\tau_1 d\tau_2
G_F(\tau_1,\tau_2) \rho(\tau_1) \cdot \rho(\tau_2) \right] 
\left. \exp \left[ \int^T_0 d\tau \rho(\tau) \cdot \gamma \right]
\right|_{\rho=0}
\nonumber 
\end{eqnarray}
which becomes
\begin{eqnarray}
\label{e21}
& = & \exp \int^T_0 d\tau \frac{1}{2} \left[ 
B^{\mu\nu}_4(\tau) \frac{\partial}{\partial \rho^{\mu}}
\frac{\partial}{\partial \rho^\nu} \right] \\
&& \times \exp \left[ - \frac{1}{2} \int^T_0 d\tau_1 d\tau_2
G_F(\tau_1,\tau_2) \rho(\tau_1) \cdot \rho(\tau_2) \right] \nonumber \\
&& \left. 
\times \exp \left[ \gamma \cdot \frac{\partial}{\partial \theta} \right]
\exp \left[ \int^T_0 d\tau \rho(\tau) \cdot \theta \right] 
\right|_{\theta = \rho= 0 } \nonumber
\end{eqnarray}
where $\theta^\mu$ is another Grassmann variable that anticommutes
with both $\rho^{\mu}(\tau)$ and $\gamma^{\mu}$.  
Under the appropriate identifications, 
we see that (\ref{e19}) can be used to reduce (\ref{e21}) to
\begin{eqnarray}
& & \hspace{-0.5cm} \P \exp \frac{1}{2}
\int^T_0 d\tau B^{\mu\nu}_4(\tau) \gamma^\mu \gamma^\nu
 = 
\Det^{1/2} \left( 
\Eins - G_F B_4
\right)
\exp \left[ \gamma \cdot \frac{\partial}{\partial \theta} \right]\\
& & \times \exp \int^T_0 d\tau_1 d\tau_2 d\tau_3
\left[ \frac{1}{2} \theta_\mu B_4^{\mu\nu}(\tau_1) 
\left\langle \tau_1 \left | \left( \Eins - G_F B_4 
\right)^{-1} \right|
\tau_2 \right\rangle_{\nu\lambda} 
G_F(\tau_2,\tau_3) B^{\lambda\sigma}_4(\tau_3)
\theta_\sigma \right] \nonumber \\
&& \left. \times \exp \left[ \int^T_0 d\tau \frac{1}{2} \theta_\mu
B^{\mu\nu}_4(\tau) \theta_\nu \right] \right|_{\theta = 0}\nonumber
\end{eqnarray}
where (using notation which is slightly different 
from that of chapter \ref{SecMaster}),
$G_F B_4$ is the Lorentz matrix-valued integral
operator with kernel
$\langle \tau_1 | G_F B_4 | \tau_2 \rangle^{\mu\nu} = G_F(\tau_1,\tau_2)
B_4^{\mu\nu}(\tau_2)$.
If we now expand $e^{\gamma \cdot \frac{\partial}{\partial \theta}}$, 
only the terms up to order
$\left(\frac{\partial}{\partial \theta}\right)^4$ 
contribute due to the Grassmann nature of the $\theta$.
We consequently find that 
\begin{eqnarray}
\label{e25}
I_4 & = & \Det^{1/2} \left( \Eins  - 
G_F B_4\right) \\
& & \times \left[ 1 + \sigma_{\mu\nu} X^{\mu\nu} +
\frac{1}{2!} \varepsilon_{\mu\nu\lambda\sigma} X^{\mu\nu} X^{\lambda\sigma}
\gamma_5 \right] \nonumber 
\end{eqnarray}
where
\begin{eqnarray}
\label{e26}
X^{\mu\nu} & = & \frac{1}{2} \int^T_0
d\tau_2 d\tau_2 d\tau_3
\left[ B^{\mu\lambda}_4(\tau_1) 
\left\langle \tau_1 \left| \left( \Eins- G_F 
B_4 \right)^{-1} \right| \tau_2 \right\rangle_{\lambda \sigma} 
G_F(\tau_2,\tau_3)
B^{\sigma\nu}_4(\tau_3) \right] \nonumber \\
& & + \int^T_0 d\tau B_4^{\mu\nu} (\tau)
\end{eqnarray}

The same procedure can be used to derive a similar decomposition
when Pauli spin matrices $\vec{\sigma}$ rather than Dirac
matrices $\gamma_\mu$ occur.  (This will prove useful in connection
with our $SO(4) = SU(2) \times SU(2)$ decomposition (\ref{e37}) below.)

If we consider
\begin{equation}
\label{e27}
I_3[\vec{A}] = \P \exp \int^T_0 d\tau \vec{A} (\tau) \cdot \vec{\sigma}
\end{equation}
we note that $2i\sigma_a = \varepsilon_{abc}\sigma_b \sigma_c$, allowing
us to write
\begin{equation}
\label{e28}
I_3 = \P \exp \frac{1}{2} \int^T_0 d\tau B^{ab}_3(\tau) \sigma_a \sigma_b
\end{equation}
where $i B^{ab}_3 (\tau) \equiv \varepsilon^{abc}A_c$.  The analogue of 
(\ref{e25}) is then
(remembering that $\vec{\sigma}$ is only a triplet)
\begin{equation}
\label{e29}
I_3 = \Det^{1/2} \left( \Eins - G_F 
B_3 \right) (1+Y)
\end{equation}
where
\begin{eqnarray}
\label{e30}
Y & = & \frac{1}{2} \left[ \sigma_a , \sigma_b \right]
\left\{ \frac{1}{2} \int^T_0 d\tau_1 d\tau_2 d\tau_3 
\left[ B_3^{ac} (\tau_1) 
\left\langle \tau_1 \left| 
\left( \Eins- G_F
B_3 \right)^{-1} \right| \tau_2 \right\rangle_{cd} G_F(\tau_2,\tau_3 )
B_3^{db} (\tau_3) \right] \right. \nonumber \\
& & \left. \hspace{0.5cm}
+ \frac{1}{2} \int^T_0 d\tau B^{ab}_3(\tau) \right\}
\end{eqnarray}
This is a generalization of the spin factor of Polyakov \cite{Polyakov}.

There is an alternate expression for $I_3$ in (\ref{e27})
that also proves useful below.  In this approach
we apply the analogue of equations (\ref{e5}) and
(\ref{e6}) to arrive at  
\begin{eqnarray}
\label{e33}
I_3[\vec{A}]
 & = & \P \exp \int^T_0 \vec{A} (\tau) \cdot \vec{\sigma} \nonumber \\
& = & \left. \Phat \exp \left[ \int^T_0 d\tau \vec{A}(\tau) \cdot 
\frac{\partial}{\partial \vec{\rho}(\tau)} \right] 
\P \exp \left[ \int^T_0 d\tau 
\vec{\rho}(\tau) \cdot \vec{\sigma} \right] 
\right|_{\vec{\rho} = 0}
\end{eqnarray}
where again, $\vec{\rho}(\tau)$ is a Grassmann source which
anticommutes with the Pauli matrices $\vec{\sigma}$.  
Then, after applying equation (\ref{e6}),
this can be written as
\begin{eqnarray}
\label{e34}
I_3 [ \vec{A} ]& = & \Phat \exp \left[ \int^T_0 d\tau \vec{A}(\tau) \cdot
\frac{\partial}{\partial \vec{\rho}(\tau)} \right] 
\\
& & \left. \times \exp \left[ - \frac{1}{2} \int^T_0
d\tau_1 d\tau_2 G_F(\tau_1,\tau_2) \vec{\rho}(\tau_1)
\cdot \vec{\rho}(\tau_2) \right]
\exp \left( T \vec{\zero \rho} \cdot \vec{\sigma} \right) 
\right|_{\vec{\rho} =0} \nonumber 
\end{eqnarray}
where $T \rho_0^a = \int^T_0 \rho^a(\tau) d\tau$.
Again, since $\vec{\zero \rho}$ is a Grassmann triplet, only terms up to
order $(\rho_0)^3$ contribute in the expansion of the last exponential
in (\ref{e34}), leaving us with%
\footnote{Care must be taken here since the assumption
that $\{ \rho , \sigma \}=0$ is inconsistent with the multiplication
rule for the $\sigma$-matrices.  Thus, it is important that the simplification
of the $SU(2)$ structure be carried out only after 
the functional
derivatives with respect to $\rho$ are allowed to act, or 
equivalently, after all $\sigma$ matrices have been 
be ordered to the right of all $\rho$ sources.} 
\begin{eqnarray}
\label{e36}
I_3 [ \vec{A} ]& = & \Phat \exp \left[ \int^T_0 d\tau \vec{A}(\tau)
\cdot \frac{\partial}{\partial \vec{\rho} (\tau) } \right] 
\exp \left[ - \frac{1}{2} \int^T_0 d\tau_1 d\tau_2 \, 
G_F(\tau_1,\tau_2) \vec{\rho} (\tau_1) \cdot
\vec{\rho}(\tau_2) \right] \nonumber \\
& & \times \left. \left[ 1 + T \vec{\zero \rho} \cdot
\vec{\sigma} 
- \frac{i \varepsilon^{abc} }{2!} 
T^2 \rho_0^a 
\rho_0^b \sigma^c
-
\frac{i}{3!} \varepsilon_{abc} T^3
\rho_0^a 
\rho_0^b 
\rho_0^c 
\right]\right|_{\vec{\rho} = 0}
\end{eqnarray}

The expressions for $I_3$ and $I_4$ are in fact related.  To see this, we
note that the 't Hooft symbols $\eta^{\pm}_{a\mu\nu}$ 
\cite{abc} can be used to write
\begin{equation}
\label{e37}
\sigma_{\mu\nu}  
= i 
\left(
\begin{array}{cc} 
\eta^+_{a\mu\nu} \sigma_a & 0 \\
0 & \eta^-_{a\mu\nu} \sigma_a 
\end{array}
\right)
\; , \;\;\; \gamma_5 = 
\left(  
\begin{array}{cc}
+1 & 0 \\ 0 & -1 
\end{array} \right)
\end{equation}
using an appropriate representation of 
the Euclidean Dirac matrices,   
and
\begin{eqnarray}
\eta^{\pm}_{aij} = \varepsilon_{aij} & \hspace{1cm} & i,j = 1,2,3 \nonumber \\
\eta^{\pm}_{a4j} = \pm \delta_{aj} & \hspace{1cm} & j = 1,2,3 \\
\eta^{\pm}_{a44} = 0 \, .\nonumber  
\end{eqnarray}
(Useful properties of the $\eta^{\pm}_{a\mu\nu}$ are given in
\cite{abc}.)  As a 
consequence, when taking the trace over spinor indices
of $K$ in (\ref{e4}) we can write

\begin{eqnarray}
\label{e39}
\tr K(y,z;T) & = & \sum_{\pm} 
\tr_\sigma \int^{x(T) = y}_{x(0)=z} \D x(\tau)
\P \exp \int^T_0 d\tau \left[ - \L_{\pm}(x,\vec{\sigma}) 
\right] \\
& = & 2 \sum_{\pm}  
\int^{x(T) = y}_{x(0)=z} \D x(\tau)
\Phat \exp \int^T_0 d\tau \left[ - \L_{\pm}\left(x,\case{\partial}{\partial 
\vec{\rho}}
\right)
d\tau \right] \label{e40} \\
& & \times \left. \exp \left[ - \frac{1}{2} \int^T_0 d\tau_1 d\tau_2 \, 
G_F(\tau_1,\tau_2)
\vec{\rho}(\tau_1) \cdot \vec{\rho}(\tau_2) \right]
\left[ 1 - \frac{i T^3}{3!} 
\rho_0^a
\rho_0^b
\rho_0^c
\varepsilon_{abc}
\right]
\right|_{\rho = 0 } \nonumber
\end{eqnarray}
where
\begin{equation}
\label{Lpm}
\L_{\pm}(x,\vec{\sigma}) \equiv \frac{ {\dot x}^2 }{4} + i {\dot x} \cdot A
+ \frac{1}{2} \sigma_a \eta^{\pm}_{a \mu \nu} F_{\mu \nu}
\pm \left( \sigma_a \eta^{\pm}_{a \mu \nu} {\dot x}^\mu A^\nu_5 
+ i \partial \cdot A_5 \right)
+ ( D-2) A_5^2 \, .
\end{equation}
Equation (\ref{e40}) follows from (\ref{e39}) by direct application
of (\ref{e27}) and (\ref{e36}). 
The functional trace of $K(T)$ can be found 
from (\ref{e39}) by setting
$z=y$ and integrating over $y$, effectively imposing periodic boundary
conditions on $x(\tau)$.  Unlike in chapter~\ref{SecMaster}, there
is no Grassmann integral to be considered. 

\section{The $VVA$ anomaly}

Any new formalism for axial vector computations must, of course,
be confronted with the axial anomaly.
In \cite{axialI} it was already shown
that the present formalism correctly yields the anomalous
divergence of the chiral current in the $VVA$ function.
Moreover, a complete cancellation 
between Feynman numerators and denominators
was found in this calculation after a suitable
integration by parts.
We will now first extend that calculation to the off-shell
case, and then rederive precisely the same integrand using
the above $SU(2)\times SU(2)$ formulation. 

\subsection{The $VVA$ anomaly from  the Master Formula}

As explained in ~\cite{axialI} it is visible already at the
path integral level that in the present formalism the
anomalous divergence implied by the ABJ anomaly \cite{abj}
is generally confined to the axial-vector currents.
Calculation of the $VV\partial\cdot A$ amplitude
either using the master formula eq.(\ref{axialoddmaster})
or by a direct Wick-contraction of eq.(\ref{repMvectorNaxial}) 
yields the following parameter integral,

\bear
k_3^{\rho}
\langle A^{\mu}A^{\nu}A_5^{\rho}\rangle
&=&
2
\varepsilon^{\mu\nu\kappa\lambda}k_1^{\kappa}k_2^{\lambda}
\Tint
{(4\pi T)}^{-2}
\prod_{i=1}^3\int_0^Td\tau_i
\non\\&&
\times \exp\Bigl\lbrack
\Bigl(G_{B12} -G_{B13}-G_{B23}\Bigr)k_1\cdot k_2 
-G_{B13}k_1^2 -G_{B23}k_2^2
\Bigr\rbrack
\non\\
&&\times
\biggl\lbrace
(k_1+k_2)^2 +
(\dot G_{B12} +\dot G_{B23} +\dot G_{B31})
(\dot G_{B13}-\dot G_{B23})
k_1\cdot k_2
-(\ddot G_{B13} +\ddot G_{B23})
\biggr\rbrace
\non\\
\label{divAAA5}
\ear
Here momentum conservation has been used to eliminate
$k_3$. 
Removing the second derivatives
$\ddot G_{B13}$ ($\ddot G_{B23}$) by a
partial integration in $\tau_1$ ($\tau_2$),
the expression in brackets turns into

\bear
&&k_1\cdot k_2
\,
\biggl\lbrace
2 -(\dot G_{B12}+\dot G_{B23}+\dot G_{B31})^2
+ \dot G_{B12}^2 - \dot G_{B13}^2 - \dot G_{B23}^2
\biggr\rbrace
+k_1^2 (1-\dot G_{B13}^2) + k_2^2 (1-\dot G_{B23}^2)
\non\\&&
=
-{4\over T} \Bigl\lbrack
\Bigl(G_{B12} -G_{B13}-G_{B23}\Bigr)k_1\cdot k_2 
-G_{B13}k_1^2 -G_{B23}k_2^2
\Bigr\rbrack
\label{rewritebraces}
\ear
In the last step we used the identities

\bear
\dot G_{Bij} + \dot G_{Bjk} + \dot G_{Bki}
&=&
-{\rm sign}(\tau_i -\tau_j)
{\rm sign}(\tau_j -\tau_k)
{\rm sign}(\tau_k -\tau_i)
\non\\
\dot G_{Bij}^2 &=& 1-4{G_{Bij}\over T}
\label{idGdot}
\ear\no
This is precisely the same expression which appears also in the
exponential factor in (\ref{divAAA5}). After 
performing the trivial $T$ - integral
we find therefore a complete cancellation between the Feynman
numerator and denominator polynomials, and obtain
without further integration the desired result,

\be
k_3^{\rho}
\langle A^{\mu}A^{\nu}A_5^{\rho}\rangle
=
{8\over {(4\pi)}^2}
\varepsilon^{\mu\nu\kappa\lambda}k_1^{\kappa}k_2^{\lambda}
\label{PCAC}
\ee

\subsection{The $VVA$ anomaly from  $SU(2)\times SU(2)$}

Alternatively we can 
compute this three-point function also
using eq.(\ref{e40}).
Evaluating the heat kernel in (\ref{e40})
with the background configuration
\begin{eqnarray}
A^\mu(x) & = & \varepsilon^\mu_1 e^{ik_1 \cdot x} 
+ \varepsilon^\mu_2 e^{ik_2 \cdot x} \nonumber \\
A_5^\mu(x) & = & \varepsilon_3^\mu e^{ik_3 \cdot x} \, ,
\end{eqnarray}
and extracting the contribution which is linear in all three
polarization vectors $\varepsilon_j^\mu$ we obtain
\begin{eqnarray}
\left.\tr \, K \right|_{VVA}& = & 2 {\sum_{\pm} } 
\int_{x(0) = x(T)}
\D x(\tau) e^{- \int^T_0 \frac{ {\dot x}^2}{4} d\tau }\Phat
\nonumber \\
& &  \times \int^T_0 d\tau_1 (-1) \left[ i {\dot x}_{1{\nu_1}}
+ i \eta^{\pm}_{a_1\mu_1\nu_1} {k_1}_{\mu_1}
\frac{\partial}{\partial \rho_{a_1}(\tau_1)} \right] \varepsilon_1^{\nu_1}
e^{i k_1 \cdot x_1} \nonumber \\
& & \times \int^T_0 d\tau_2 (-1) \left[ i {\dot x}_{2{\nu_2}}
+ i \eta^{\pm}_{a_2\mu_2\nu_2} {k_2}_{\mu_2}
\frac{\partial}{\partial \rho_{a_2}(\tau_2)} \right ] \varepsilon_2^{\nu_2}
e^{i k_2 \cdot x_2} \\
& & \times (\pm) \int^T_0 d\tau_3 (-1) \left[ -{k_3}_{\nu_2} 
+ \eta^{\pm}_{a_3{\mu_3}\nu_3} {\dot x}_{3{\mu_3}}
\frac{\partial}{\partial \rho_{a_3}(\tau_3)} \right]
\varepsilon_3^{\nu_3} e^{ik_3 \cdot x_3}  \nonumber \\
& & \left. \times \exp \left[ - \frac{1}{2} \int^T_0 d\tau_1 d\tau_2
\, G_F(\tau_1,\tau_2)
\vec{\rho}(\tau_1) \cdot \vec{\rho}(\tau_2) \right]
\left[ 1 - \frac{i T^3}{3!}
\rho_0^a
\rho_0^b
\rho_0^c
\varepsilon_{abc}
\right]
\right|_{\rho = 0 } \nonumber \\
& = & - 2 
\varepsilon_{1{\nu_1}} 
\varepsilon_{2{\nu_2}} 
\varepsilon_{3{\nu_3}} 
{\sum_{\pm} } (\pm) \int_{x(0) = x(T) }
\D x(\tau) e^{- \int^T_0 \frac{ {\dot x}^2}{4} d\tau } 
\int^T_0 d\tau_1 d\tau_2 d\tau_3 e^{i \left(k_1 \cdot x_1
+ k_2 \cdot x_2 + k_3 \cdot x_3 \right) } \nonumber \\
& & 
\times \left[ 
{\dot x}_{1\nu_1} {\dot x}_{2\nu_2} k_{3\nu_3}
- \eta^{\pm}_{a \mu_1 \nu_1} k_{1\mu_1}  \eta^{\pm}_{a \mu_2 \nu_2 }
k_{2 \mu_2} 
G_F(\tau_2,\tau_1) 
\sign(\tau_1 - \tau_2) 
k_{3\nu_3} 
\right.
 \label{VVA}\\
& & \hspace{0.5cm}
+ {\dot x}_{1\nu_1} \eta^{\pm}_{a\mu_2\nu_2}
k_{2\mu_2} \eta^{\pm}_{a\mu_3\nu_3}
{\dot x}_{3 \mu_3} G_F(\tau_3,\tau_2) \sign(\tau_2-\tau_3)
+ (1 \leftrightarrow 2) \nonumber \\  
& & \hspace{0.5cm}
\left.
+ \eta^{\pm}_{a \mu_1 \nu_1} k_{1\mu_1} \eta^{\pm}_{b \mu_2 \nu_2}
k_{2\mu_2} \eta^{\pm}_{c \mu_3 \nu_3} {\dot x}_{3 \mu_3 }
i \varepsilon_{cba} \sign(\tau_1, \tau_2, \tau_3) \right]
\nonumber
\end{eqnarray}
(Note the simplification 
\mbox{$G_F(\tau,\tau') \sign(\tau'-\tau) = -1$}.)
Here the $\Phat$ ordering of anticommuting variables has been
resolved using the usual signum function as well as
its following higher-point generalization,

\bear
\sign(\tau_1,\tau_2,\ldots,\tau_n) &\equiv& 
\prod_{1\le i<j\le n}
\sign(\tau_i -\tau_j) \, .
\label{defsign123}
\ear\no
Further simplification is provided by 
the algebraic properties
of the $\eta$ symbols \cite{abc}.
Of particular use are the identities
\begin{eqnarray}
\eta^{\pm}_{a \mu_1 \nu_1} \eta^\pm_{a \mu_2 \nu_2}
&  = & \delta_{\mu_1 \mu_2} \delta_{\nu_1 \nu_2}
- \delta_{\mu_1 \nu_2} \delta_{\nu_1 \mu_2}
\mp \varepsilon_{\mu_1 \nu_1 \mu_2 \nu_2 } \\
\varepsilon_{abc} \eta^\pm_{b \mu_2 \nu_2}
\eta^\pm_{c \mu_3 \nu_3}
& = & 
\delta_{\mu_2 \mu_3} \eta^\pm_{a \nu_2 \nu_3}
- \delta_{\mu_2 \nu_3} \eta^\pm_{a \nu_2 \mu_3}
- \delta_{\nu_2 \mu_3} \eta^\pm_{a \mu_2 \nu_3}
+ \delta_{\nu_2 \nu_3} \eta^\pm_{a \mu_2 \mu_3} \nonumber
\end{eqnarray}
The application of these rules to an equation
such as (\ref{VVA}) can be greatly simplified
if one recognizes that the rules for $\eta^+_{a\mu\nu}$
differ from those for $\eta^-_{a\mu\nu}$ only in the sign coefficient
of $\varepsilon_{\mu\nu\rho\lambda}$.  The effect of subtracting the $+$ and $-$
contributions is to double those terms which are proportional to  
the $\varepsilon$ - tensor 
while canceling those which are not.
Furthermore, for the ABJ anomaly, we are interested
in the longitudinal axial polarization
$\varepsilon_3 \propto  k_3$ which together with
conservation of momentum eliminates terms of the form
$\varepsilon_{\mu_1\mu_2\nu_3\lambda} k_{1\mu_1} k_{2\mu_2} k_{3\nu_3}$.

We then integrate over the 
trajectories $x(\tau)$ using same contraction rules and
path integral normalization as in chapter~\ref{SecMaster}.
This yields
\begin{eqnarray}
k_{3\nu_3} \left\langle A_{\nu_1}(k_1) A_{\nu_2}(k_2) A_{5\nu_3}(k_3) 
\right\rangle
& = & 2 \varepsilon_{\nu_1 \nu_2 \mu_1 \mu_2} k_1^{\mu_1} k_2^{\mu_2}
\int^\infty_0 \frac{dT}{T} (4 \pi T)^{-2} \int_0^T d\tau_1 d\tau_2
d\tau_3 \nonumber \\ 
& & \times \exp \left[ 
k_1 \cdot k_2 G_{B12} 
+ k_2 \cdot k_3 G_{B23} 
+ k_1 \cdot k_3 G_{B13} \right]
 \\
& & \times \left\{ k_3^2 + \left( \dot G_{B13} - \dot G_{B23} \right)
\sign(\tau_1, \tau_2, \tau_3 ) k_1 \cdot k_2 
- \left( \ddot G_{B13} + \ddot G_{B23} \right)
\right\} \nonumber 
\end{eqnarray} 
Using eq.(\ref{idGdot}) we see that
this is precisely the same integrand as was obtained above in
eq.(\ref{divAAA5}). 

\section{Inclusion of external Fermions}

The heat kernel evaluated by using the quantum
mechanical path integral in conjunction with the FG approach to replacing
Dirac matrices by a path integral over Grassmann variables can also
be employed to compute Green's functions involving external
fermions and quantized gauge fields.
This we illustrate by considering the Lagrangian
\begin{equation}
\label{e100}
L = \overline \psi \left( p \!\! / + A \!\!\!/
+ \gamma_5 A \!\!\! /_5 \right) \psi
- \frac{1}{2} \left( \partial_\mu A_{5\nu} \right) \left( \partial^\mu
A_5^\nu \right) - \frac{1}{2} \left( \partial_\mu A_\nu \right)
\left( \partial^\mu A^\nu \right)
\end{equation}
in the Feynman gauge for an axial vector field $A_{5\mu}$, a vector
field $A_\mu$ and a massless
spinor field $\psi$.  If we have an external
component to the spinor field $\psi$, then the one-loop effective
action involves the functional determinant.
\begin{equation}
\label{e101}
\sdet^{-1/2} M = \sdet^{-1/2} \left(
\begin{array}{cccc}
0 & p \!\! / & \gamma_\nu & \gamma_5 \gamma_\nu \psi \\
- p \!\! / ^ T & 0 & - \left( \overline \psi \gamma_\nu \right)^T &
- \left( \overline \psi \gamma_5 \gamma_\nu \right) ^T \\
\left( \gamma_\mu \psi \right)^T & \overline \psi \gamma_\mu &
- p^2 & 0 \\
\left( \gamma_5 \gamma_\mu \psi \right) ^ T &
\overline \psi \gamma_5 \gamma_\mu & 0 & -p^2
\end{array}
\right) \, .
\end{equation}
Premultiplying (\ref{e101}) by
\begin{equation}
\label{e102}
X = \left( 
\begin{array}{cccc}
0 & - p \!\! / ^ T & 0 & 0 \\
p \!\! / & 0 & 0 & 0 \\
0 & 0 & -1 & 0 \\
0 & 0 & 0 & -1 
\end{array}
\right)
\end{equation}
leads to
\begin{eqnarray}
\label{e103}
\sdet^{-1/2} XM & = & \sdet^{-1/2} \left(
\begin{array}{cccc}
p^2 & 0 & \left( \overline \psi \gamma_\nu p \!\! / \right) ^ T &
\left( \overline \psi \gamma_5 \gamma_\nu p \!\! / \right) ^T \\
0 & p^2 & p \!\! / \gamma_\nu \psi & p \!\! / \gamma_5 \gamma_\nu
\psi \\
- \left( \gamma_\mu \psi \right)^ T & - \overline \psi \gamma_\mu &
p^2 & 0 \\
- \left( \gamma_5 \gamma_\mu \psi \right) ^ T & 
- \overline \psi \gamma_5 \gamma_\mu & 0 & p^2 
\end{array}
\right) \\
& = &
\sdet^{-1/2} \left\{ \left[ \left(
\begin{array}{cccc}
p_\rho & 0 & 0 & 0 \\
0 & p_\rho & 0 & 0 \\
0 & 0 & p_\rho & 0 \\
0 & 0 & 0 & p_\rho
\end{array}
\right)
+ \frac{1}{2}
\left(
\begin{array}{cccc}
0 & 0 & \left( \overline \psi \gamma_\nu \gamma_\rho \right) ^ T &
\left( \overline \psi \gamma_5 \gamma_\nu \gamma_\rho \right) ^ T \\
0 & 0 & \gamma_\rho \gamma_\nu \psi & \gamma_\rho \gamma_5 \gamma_\nu
\psi \\
0 & 0 & 0 & 0 \\
0 & 0 & 0 & 0 
\end{array}
\right)
\right] ^ 2 \right. \nonumber \\
& & \hspace{0.5cm}
+ \frac{1}{2} \left(
\begin{array}{cccc}
0 & 0 & i \left( \overline \psi \gamma_\nu \gamma_\rho \right)^T_{,\rho}
& i \left( \overline \psi \gamma_5 \gamma_\nu \gamma_\rho \right)^T_{,\rho} \\
0 & 0 & -i \gamma_\rho \gamma_\nu \psi_{,\rho} &
-i \gamma_\rho \gamma_5 \gamma_\nu \psi_{,\rho} \\
0 & 0 & 0 & 0 \\
0 & 0 & 0 & 0
\end{array}
\right) \nonumber \\ 
& & \hspace{0.5cm} + \left. \left(
\begin{array}{cccc}
0 & 0 & 0 & 0 \\
0 & 0 & 0 & 0 \\
- \left( \gamma_\mu \psi \right)^T & - \overline \psi \gamma_\mu & 0 & 0 \\
- \left( \gamma_5 \gamma_\mu \psi \right)^ T & - \overline \psi \gamma_5
\gamma_\mu & 0 & 0
\end{array}
\right)
\right\} 
\, . \nonumber
\end{eqnarray}
This is of the form
\begin{equation}
\sdet^{-1/2} \left[ (p-A)^2 + V \right]
\end{equation}
provided
\begin{equation}
\label{e104}
A_\rho = - \frac{1}{2}
\left(
\begin{array}{cccc}
0 & 0 & \left( \overline \psi \gamma_\nu \gamma_\rho \right) ^ T &
\left( \overline \psi \gamma_5 \gamma_\nu \gamma_\rho \right) ^ T \\
0 & 0 & \gamma_\rho \gamma_\nu \psi & \gamma_\rho \gamma_5 \gamma_\nu
\psi \\
0 & 0 & 0 & 0 \\
0 & 0 & 0 & 0
\end{array}
\right)
\end{equation}
and
\begin{equation}
\label{e105}
V = \left(
\begin{array}{cccc}
0 & 0 & \frac{i}{2} \left( \overline 
\psi \gamma_\nu \gamma_\rho \right)^T_{,\rho}
& \frac{i}{2} \left( \overline
 \psi \gamma_5 \gamma_\nu \gamma_\rho \right)^T_{,\rho} \\
0 & 0 & -\frac{i}{2} \gamma_\rho \gamma_\nu \psi_{,\rho} &
-\frac{i}{2} \gamma_\rho \gamma_5 \gamma_\nu \psi_{,\rho} \\
- \left( \gamma_\mu \psi \right)^T & - \overline \psi \gamma_\mu & 0 & 0 \\
- \left( \gamma_5 \gamma_\mu \psi \right)^ T & - \overline \psi \gamma_5
\gamma_\mu & 0 & 0
\end{array}
\right)
\end{equation}
Since we can express
\begin{eqnarray}
\label{e106}
\ln \sdet^{-1/2} \left[ (p-A)^2 + V \right] & = &
- \half \str \ln \left[ (p-A)^2 + V \right] \\
& = & \frac{1}{2 \Gamma(s) }
\int^\infty_0 dT \, T^{s-1} \str \left\langle
x \left| \exp -T \left[ (p-A)^2 + V \right] \right|
y \right\rangle \nonumber
\end{eqnarray}
where operator regularization \cite{OR} has been used
\footnote{Of course one could as well use heat kernel
regularisation.}, we see that the
one-loop effective potential can be worked out by considering
the path integral for the matrix in (\ref{e106})
\begin{equation}
\label{e107}
M_{yz} = \P \int^{x(T) = y}_{x(0) = z} \D x (\tau) 
 \exp \int^T_0 d\tau \left[
- \frac{ \dot{x}^2 }{4} + i A(x(\tau)) \cdot \dot{x}(\tau) -
V(x(\tau)) \right] \, .
\end{equation}
The FG procedure for replacing the $\gamma$ matrices
occurring in (\ref{e104}) and (\ref{e105}) when they are inserted
into (\ref{e107}) is now straightforward; an example of how this is done in
the context of the Yukawa model is given in
\cite{TonyMe}.

\section{Discussion}

In this paper we have considered various aspects
of handling $\gamma$-matrices that appear in the quantum mechanical
path integral which arises when computing radiative
effects in models in which a spinor is coupled to both
a vector and axial vector field.  These $\gamma$-matrices
can either be replaced by an integral over a Grassmann parameter, or
reduced to a ``spin factor''.

The Dirac matrix valued worldline Lagrangian 
in (\ref{e4}) 
(and, indeed
the squared operator which appears in equation (\ref{e2}))
commutes with $\gamma_5$ and leads to an apparent
decoupling of the left and right-handed spin sectors,
as we saw in equation (\ref{e39}).  
This property also led to simplifications in
the calculation of the axial anomaly. 
In models with isotopic spin symmetry,
$SU(2)$ matrices $\tau_a$ enter into the path integral
and can be treated in a manner identical to the
way in which the Pauli spin 
matrices $\sigma_a$ have been treated in (\ref{e33})-(\ref{e34}).
In fact, generators of the spinor representation of a group 
$SO(N)$ can be treated in a manner similar to 
equations~(\ref{e4})-(\ref{L}).  
This possibility was also considered in \cite{GitmanSvartsman}.

On the applied side, our main result is in the master formulas
(\ref{axialevenmaster}),(\ref{axialoddmaster}). They constitute
a generalization of the QED Bern-Kosower master formula for
photon scattering. To discuss the potential
usefulness of these formulas it is necessary to restate some
known facts about the relation between the ``string-inspired
formalism'' and standard field theory 
\cite{BernKosower,berdun,strassler}. Generally, the
parameter integrals obtained in string-inspired approaches
correspond to a second order formulation of the
field theory in question. For example, in scalar QED
the parameter integral obtained for the $N$ - photon
amplitude from the Bern-Kosower master formula
can be directly identified with the corresponding
Feynman parameter integral. This requires the rewriting
of Feynman parameters as differences of proper-times,
and the use of  momentum conservation. The seagull vertex
present in scalar QED then corresponds precisely to
the $\delta$ - function appearing in the second derivative of
the bosonic worldline Green's function. 

The same identification works also for spinor QED, however
not in its usual first-order formulation, but in a second-order
formulation. The existence of such a reformulation
of spinor QED has been known for a long time ~\cite{feygel,hostler}.
However it has rarely been applied in practice, and 
was worked out in detail only recently in
~\cite{morgan}, where also the non-abelian generalization
was investigated.
Those second-order Feynman rules, as well as their
generalization to the vector-axialvector case,
can be directly read off eq.(\ref{e2}) and are given
in appendix \ref{secondorderrules}. For the QED case they
are an extension of the scalar QED rules, with a scalar
propagator for the electron, and an additional vertex involving
the $\sigma_{\mu\nu}$ - matrix. (This is also the reason why
in QED applications the ``string-inspired'' formalism generally
allows one to obtain scalar QED results as a byproduct
of spinor QED calculations \cite{ss3,rescsc}).

Despite of this direct correspondence,
the worldline representation of the parameter integrals has
two advantages over the Feynman parameter representation.
Firstly, by rewriting the integrand in terms of the ``circular''
functions $G_B,G_F$ it is brought into a form where one can
perform partial integrations in parameter space without
generating boundary terms (in the abelian case considered here).
In the QED case this can be used to remove all $\ddot G_B$'s
appearing in the initial integrand, and to arrive at
a Feynman numerator which is completely expressed in terms of
the various $\dot G_{Bij}$, and 
homogeneous in the external momenta
\cite{BernKosower,strassler}. This rearrangement leads to a canonical
gauge invariant decomposition of the $N$ - photon amplitude
\cite{menphoton} and also to intriguing cancellations in the
calculation of the QED 2-loop $\beta$ - function \cite{ss3}.
Moreover it makes the worldline supersymmetry, implicit in the
path integral, manifest at the parameter integral level
\cite{BernKosower,strassler,menphoton}
(for more on this point see appendix C).
It will be interesting to work out the 
consequences of the partial integration
procedure for the generalized master formulas.
As we have already seen in chapter 4 it
leads to a trivialization of the parameter integrals in the
case of the ABJ anomaly. 

A second advantage of the worldline representation is
that the integrands are valid for an arbitrary ordering of the
external legs, so that one can write down whole one-loop
amplitudes ``in one piece'', with no necessity to sum over
crossed diagrams. This property is less relevant at the one-loop
level but leads to interesting consequences after
multiloop generalization \cite{ss2,ss3}.

Somewhat unusual is also the behaviour of the present
formalism with respect to chiral symmetry in the
dimensional continuation. It should be remembered
that, in field theory, one has essentially a choice between
two evils. If one preserves the anticommutation
relation between $\gamma_5$ and the other Dirac matrices
\cite{chfuhi}
then the chiral symmetry is preserved for parity-even
fermion loops, but Dirac traces with an odd number
of $\gamma_5$'s are not unambiguously defined in general,
requiring additional prescriptions. The main alternative
is to use the
't Hooft-Veltman-Breitenlohner-Maison prescription
\cite{HV,Bremai}. In this case there are no ambiguities, but
the chiral symmetry is explicitly broken, so that 
in chiral gauge theories finite renormalizations
generally become
necessary to avoid violations of the gauge Ward identities
\cite{bonneau,megamma5,trueman}.

Since our path integral representation was derived using an
anticommuting $\gamma_5$, we have not broken the chiral
symmetry. In particular, in the massless case the amplitude
with an even number of axialvectors 
should coincide with the
corresponding vector amplitude, and we have explicitly
verified this fact for the two-point case. 
(Even though the structure of the resulting Feynman numerators is quite
different from the equivalent ones derived from the ordinary
Bern-Kosower master formula.)
Nevertheless, we did not encounter any ambiguities 
even in the parity-odd case, not even in the anomaly
calculation. As was explained already in part I, 
it is clear already from the structure of the vertex
operators that only the axial vector currents
can develop anomalous divergences.
This property makes us optimistic about
the potential usefulness of this formalism
for calculations in chiral gauge theories.

Finally, let us mention some generalizations of the
formalism developed here which would be entirely
straightforward.
First, the manipulations leading to eq.(\ref{GammaAA5})
could as well be carried out with an additional scalar
background field (though in the massive case not with a pseudoscalar
one). Then, also our restriction to the abelian case
was largely a matter of expedience. Finally, the master formulas
(\ref{axialevenmaster}),(\ref{axialoddmaster})
could be easily modified to take an additional
external constant electromagnetic field
into account along the lines of \cite{rescsc}. The special case of
the 2-point vector --  axial-vector amplitude in a 
constant field was already considered
in \cite{sandansky,ioasch}.  
In addition, one could also deal with the Lorentz violating theories 
considered in~\cite{r61}.

\vspace{15pt}\noindent
{\bf Acknowledgements:}
F.A.D. would like to thank Dmitri Gitman for pointing out 
reference~\cite{Gitman98}. 
C.S. thanks M. Reuter for various suggestions, and the
Institute for Advanced Study, Princeton, for hospitality
during the final stage of this work.  We would like to thank
the authors of \cite{r61}, and \cite{Italian} for having brought those
references to our attention.
NSERC of Canada provided financial support.

\vfill\eject
\appendix
\section{Coherent state representation of the effective
action}
\label{coherentstates}

In this appendix we
use the coherent state formalism developed in
\cite{Kashiwa} to transform the functional trace 
of the heat kernel $K(x,y;T)$ in eq.(\ref{e4}) 
into a quantum mechanical
path integral. Our treatment closely parallels the one in
\cite{DHokerGagne}
except that they work in six dimensions.
Define matrices $a_r^+$ and $a_r^-$, $r=1,2$, by

\be
a_1^{\pm} = \half(\gamma_1\pm i\gamma_3),
\quad
a_2^{\pm} = \half(\gamma_2\pm i\gamma_4),
\label{defa12}
\ee\no
Those satisfy Fermi-Dirac anticommutation rules

\be
\lbrace a_r^+,a_s^-\rbrace = \delta_{rs},
\quad
\lbrace a_r^+,a_s^+\rbrace = 
\lbrace a_r^-,a_s^-\rbrace = 
0
\label{arsanticomm}
\ee
Thus we can use $a_r^+$ and $a_r^-$ as creation and
annihilation operators for a Hilbert space with a
vacuum defined by

\be
a_r^- |0\rangle =
\leftvac a_r^+ =0
\label{defavacuum}
\ee\no
Next we introduce Grassmann variables $\eta_r$ and $\bar\eta_r$,
$r=1,2$,
which anticommute with one another and with the operators
$a_r^{\pm}$, and commute with the vacuum $\rightvac$.
The coherent states are then defined as

\bear
\langle\eta\mid\equiv i\langle0\mid
(\eta_1-a_1^-)
(\eta_2-a_2^-)
\qquad
&&
\mid\eta\rangle
\equiv
\exp (-\eta_1a_1^+ -\eta_2a_2^+)
\rightvac
\non\\
\langle\bar\eta\mid
\equiv
\leftvac
\exp (-a_1^-\bar\eta_1-a_2^-\bar\eta_2)
\qquad &&
\mid\bar\eta\rangle
\equiv
i
(\bar\eta_1-a_1^+)
(\bar\eta_2-a_2^+)
\rightvac
\label{defcoherentstates}
\ear
It is easily verified that those satisfy the defining
equations for coherent states,

\bear
\langle\eta\mid a_r^- = \langle\eta\mid\eta_r
\qquad
a_r^-\mid\eta\rangle = \eta_r\mid\eta\rangle
\qquad
&&
\langle\eta\mid\bar\eta\rangle
=
\exp
(\eta_1\bar\eta_1
+\eta_2\bar\eta_2
)
\non\\
\langle\bar \eta\mid a_r^+ = \langle\bar \eta\mid\bar\eta_r
\qquad
a_r^+ \mid\bar\eta\rangle = \bar\eta_r\mid\bar\eta\rangle
\qquad
&&
\langle\bar\eta\mid\eta\rangle
=
\exp
(\bar\eta_1\eta_1
+\bar\eta_2\eta_2
)
\label{verifycoherentstates}
\ear
Also one introduces the corresponding Grassmann integrals,
defined by

\be
\int\eta_i d\eta_i
=
\int\bar\eta_i d\bar\eta_i
=i
\label{Grassmannetaint}
\ee
The $d\eta_r,d\bar\eta_r$ commute with one another and with the vacuum,
and anticommute with all Grassmann variables and the $a^{\pm}_r$.
This leads to the following completeness relation

\be
\Eins = i
\int\mid\eta\rangle\langle\eta\mid
d^2\eta
= -i
\int d^2\bar\eta
\mid\bar\eta\rangle
\langle
\bar\eta\mid
\label{completenessrelations}
\ee
($d^2\eta=d\eta_2 d\eta_1,
\,\,
d^2\bar\eta
=d\bar\eta_1
d\bar\eta_2
$), and to the following representation for a 
trace in the Fock space generated by the
$a^{\pm}_r$,

\be
\Tr (U)
= i
\int
d^2\eta\,
\langle
-\eta
\mid
U\mid
\eta\rangle
\label{Diractracerepresentation}
\ee
We can now apply these fermionic
coherent states together with the usual complete
sets of coordinate states to rewrite the
functional trace of 
(\ref{e4}) in the following way, 

\bear
\Tr
\e^{-T\Sigma}
&=& i
\int d^4x \int d^2\eta\,
\langle
x,-\eta\mid
\e^{-T\Sigma}
\mid
x,\eta\rangle
\non\\
&=& i^N
\int
\prod_{i=1}^N
\Bigl(
d^4x^id^2\eta^i
\langle
x^i,\eta^i\mid
\e^{-{T\over N}\Sigma}
\mid
x^{i+1},\eta^{i+1}
\rangle
\Bigr)
\label{rewritefunctionaltrace}
\ear
where 
$\Sigma = - (\partial_\mu 
+ i{\cal A}_\mu)^2 + V
$.
The boundary conditions on the $x$ and $\eta$ integrations
are
$(x^{N+1},\eta^{N+1})
=
(x^1,-\eta^1)$.
For the evaluation of this matrix element it will be useful to look
first at the matrix elements of elementary products of
Dirac matrices. For the product of two $\gamma$'s one finds

\be
\langle
\eta^i
\mid
\gamma_{\mu}\gamma_{\nu}
\mid
\eta^{i+1}
\rangle
=
- i
\int d^2\bar\eta^{i,i+1}
\langle
\eta^i\mid
\bar\eta^{i,i+1}
\rangle
\langle
\bar\eta^{i,i+1}
\mid\eta^{i+1}
\rangle
2
\phantom{,}^i\psi_{\mu}
\psi_{\nu}^{i+1}
,\quad
\mu\ne \nu
\label{gammatopsi}
\ee
where

\bear
\psi_{1,2}^{i+1}
\equiv
{1\over\sqrt 2}
(\eta_{1,2}^{i+1}+\bar\eta_{1,2}^{i,i+1})
,\qquad
&&
\psi_{3,4}^{i+1}
\equiv
{i\over\sqrt 2}
(\eta_{1,2}^{i+1}
-\bar\eta_{1,2}^{i,i+1}
)
\non\\
\phantom{,}^i\psi_{1,2}
\equiv
{1\over\sqrt 2}
(\eta_{1,2}^i
+
\bar\eta_{1,2}^{i,i+1}
),\qquad
&&
\phantom{,}^i
\psi_{3,4}
\equiv
{i\over\sqrt 2}
(\eta_{1,2}^i -
\bar\eta_{1,2}^{i,i+1}
)
\label{etatopsi}
\ear
To verify this equation one
rewrites the Dirac matrices
in terms of the $a_r^{\pm}$
and then
inserts a complete set of coherent
states
$\mid\bar\eta^{i,i+1}\rangle$
in between them. 

To take also the $\gamma_5$ - matrix into account it is
crucial to observe that, expressed in terms
of the $a_r^{\pm}$, it is identical to the fermion number
counter or ``G-parity operator''
${(-1)}^F$ \cite{Gaume},

\be
\gamma_5= {(-1)}^F = (1-2F_1)(1-2F_2)
\label{gamma5=fnco}
\ee
where 

\be
F \equiv F_1 +F_2, \quad {\rm with} \quad F_i = a_i^+ a_i^-
\label{defFi}
\ee
{}From the identity

\be
\langle -\eta\mid{(-)}^F
=i\leftvac\prod_{r=1}^2
(-\eta_r-a_r^-)(1-2a_r^+a_r^-)
=i
\leftvac
\prod_{r=1}^2
(-\eta_r+a_r^-)
=\langle\eta\mid
\label{fncoaction}
\ee
it is clear that the presence of
${(-1)}^F$ can be taken into account
by switching the boundary conditions 
on the Grassmann path integral from antiperiodic
to periodic.

\no
With this information it is now easy to compute that

\bear
\langle
x^i,\eta^i\mid
\e^{-T\Sigma[p,A,A_5,\gamma_{\mu}\gamma_{\nu},\gamma_5]}
\mid x^{i+1},\eta^{i+1}\rangle
&=&
-
{i\over {(2\pi)}^4}
\int
d^4p^{i,i+1}
d^2\bar\eta^{i,i+1}
\e^{i(x^i-x^{i+1})p^{i+1}
+(\eta^i-\eta^{i+1})_r
\bar\eta_r^{i,i+1}}
\non\\
&&\!\!\!\!\!\!\!\!\!\!\!\Mneg\Mneg
\times
\Bigl[
1-{T\over N}
\Sigma[p^{i,i+1},A^{i,i+1},A_5^{i,i+1},
2\phantom{,}^i\psi_{\mu}
\psi_{\nu}^{i+1},{(-1)}^F]
+O\bigl({T^2\over N^2}\bigr)
\Bigr]
\label{matrixelement}
\ear
Here the superscript $\phi^{i,i+1}$ on a field denotes
the average of the corresponding fields with superscripts
$i$ and $i+1$. 
Inserting this result back into eq.(\ref{rewritefunctionaltrace})
one obtains, after symmetrizing the positions of the Grassmann
variables in the exponentials,

\bear
\Tr
\e^{-T\Sigma}
&=&
{1\over {(2\pi)}^{4N}}
\int
\prod_{i=1}^N
d^4x^i
d^4p^{i,i+1}
d^2\eta^i
d^2\bar\eta^{i,i+1}
\Bigl(
1-{T\over N}
\Sigma_i
+
{\rm O}
({T^2\over N^2})
\Bigr)
\non\\
&&\times
\exp
\biggl\lbrace
\sum_{i=1}^N
\Bigl[
i(x^i-x^{i+1})
p^{i,i+1}
+\half
(\eta_r^i
-\eta_r^{i+1})
\bar\eta_r^{i,i+1}
-\half
\eta_r^i
(\bar\eta_r^{i-1,i}-\bar\eta_r^{i,i+1})
\Bigr]
\biggr\rbrace
\label{Tracediscretized}
\ear
Introducing an interpolating proper-time
$\tau$ such that $\tau_1=T$, $\tau^{N+1}=0$,
and $\tau^i-\tau^{i+1}  = {T\over N}$, and
taking the limit $N\to\infty$
in the usual naive way, we finally obtain the
following path integral representation,

\be
\Tr
\e^{-T\Sigma}
=
\int{\cal D}p
\int{\cal D}x
\int_P{\cal D}\eta{\cal D}\bar\eta
\exp
\biggl\lbrace
\int_0^Td\tau\,
\Bigl[
i\dot x\cdot p
+\half
\dot\eta_r\bar\eta_r
-\half
\eta_r
\dot{\bar\eta_r}
-\Sigma [p,A,A_5,2\psi_{\mu}\psi_{\nu},\hat\gamma_5]
\Bigr]
\biggr\rbrace
\label{pxpi}
\ee
The ``P'' denotes the antiperiodic boundary conditions which we
have for $\eta,\bar\eta$ originally.
The operator ${(-1)}^F$ has turned into an operator $\hat\gamma_5$
whose only raison d' {\^e}tre is to determine the boundary conditions
of the Grassmann path integral; after expansion of the interaction
exponential a given term will have to be
evaluated using antiperiodic (periodic) boundary conditions
on ${\cal D}\eta$, ${\cal D}\bar\eta$, if it contains
$\hat\gamma_5$ at an even (odd) power. After the boundary conditions
are determined $\hat\gamma_5$ can be replaced by unity.

\no 
The continuum limit of eqs.(\ref{etatopsi})
is

\be
\psi_{1,2}(\tau)
=
{1\over\sqrt 2}
(\eta_{1,2}(\tau)+\bar\eta_{1,2}(\tau)),
\qquad
\psi_{3,4}(\tau)
=
{i\over\sqrt 2}
(\eta_{1,2}(\tau)-\bar\eta_{1,2}(\tau))
\label{etatopsicont}
\ee
This suggests a change of variables from
$\eta,\bar\eta$ to $\psi$, which we
complete by rewriting the fermionic kinetic term,

\be
\half
\dot\eta_r\bar\eta_r
-\half
\eta_r\dot{\bar\eta}_r
=
-\half
\psi_{\mu}\dot\psi_{\mu}
\label{etatopsikin}
\ee
The boundary conditions are now 
$(x(T),\psi(T))=(x(0),-\psi(0))$.
Finally, we note that the momentum path integral
is Gaussian, and perform it by a naive completion
of the square (for a less unscrupulous treatment 
of this point see again \cite{DHokerGagne}, 
as well as for the various normalization factors involved). 
This brings us to our following
final result
\footnote{Our definition of the Euclidean effective action
differs by a sign from the one used in \cite{DHokerGagne}.}

\bear
\Gamma_{\rm spin}[A,A_5]
&=& -2\, \int_0^\infty \, 
\frac{dT}{T} 
\e^{-m^2T}
\Dx
\Dpsi
\, \e^
{
-\int_0^Td\tau\,
L(\tau)
}\label{vapi}\\
L(\tau) &=&
\kinb
+\half\psi_{\mu}\dot\psi^{\mu}
+i\dot x^{\mu}A_{\mu}
-i\psi^{\mu}F_{\mu\nu}\psi^{\nu}
-2i\hat\gamma_5\dot x^{\mu}\psi_{\mu}\psi_{\nu}A_5^{\nu}
+i\hat\gamma_5\partial_{\mu}A^{\mu}_5
+(D-2)A_5^2
\non
\ear\no

\section{Wick's theorem and the spin path integral}
\label{NoPI}

In this appendix we discuss the connection between Wick's
theorem (\ref{e6}) as it is used in (\ref{e7})
and the Feynman rules of the spin path integral
(\ref{GammaAA5})
as described in 
chapter~\ref{SecMaster}, 
where one is generally interested in the 
trace of the heat kernel
$$
\int d^Dy \, \tr \, K(y,y;T) \, .
$$

Expanding (\ref{e7}) in powers of its axialvector (and, in the more 
general case, 
pseudoscalar) couplings (i.e in powers of $\gamma_5$), 
we notice that there are only two possible nonzero
traces which can arise:

For {\em even} powers in the $\gamma_5$ expansion we have 
\begin{equation}
\tr \exp \left(  T \zero \rho \cdot \gamma \right) = 4 \, .
\end{equation}
In this case, the partial derivatives $\frac{\partial}{\partial \rho}$
which occur in (\ref{e7}) are forced to act on the propagator 
exponential and lead to Feynman contractions which resemble
(\ref{modcorrelators})
$$
\langle \psi^\mu(\tau_1) \psi^\nu(\tau_2) 
\rangle = \frac{1}{2} \sign(\tau_1 - \tau_2)
g^{\mu\nu}
$$
where we make the correspondence
$\frac{\partial}{\partial \rho^\mu(\tau)} \rightarrow 
\sqrt{2} \psi^\mu(\tau)$.

For {\em odd} powers we have
\begin{equation}
\tr \, \gamma_5 \exp \left( T \zero \rho \cdot \gamma \right) = 
\frac{4 T^4}{4!} \varepsilon^{\mu\nu\lambda\rho} 
\zero \rho^\mu 
\zero \rho^\nu 
\zero \rho^\lambda 
\zero \rho^\rho \, .
\end{equation}
For such processes, the zero-mode
contributions to the propagator exponent are prohibited from contributing,
due to the Grassmann character of $\zero \rho$.  As a result, we may
safely remove the zero modes from the exponential,  
replacing it with
\begin{equation}
\label{nonzeroprop}
\exp \left[ - \frac{1}{2} \int^T_0 d\tau_1 d\tau_2
\left( \sign(\tau_1 - \tau_2) - 2\frac{\tau_1 - \tau_2}{T} \right)
\rho(\tau_1) \cdot \rho(\tau_2)
\right] \, .
\end{equation}
So that the partial derivatives $\frac{\partial}{\partial \rho}$
may act either on the zero mode terms
or the propagator exponential (\ref{nonzeroprop}).
Making the correspondence 
$\frac{\partial}{\partial \rho^\mu (\tau)}
\rightarrow \sqrt{2} \left[ \psi_0^\mu + \xi^\mu(\tau)\right]$
we arrive at 
a Feynman rule similar to the zero-mode rule in 
(\ref{zeromodeintegral})
$$
\int d^4 \psi_0 \, \psi_0^\mu \psi_0^\nu \psi_0^\kappa \psi_0^\lambda
= \varepsilon^{\mu\nu\kappa\lambda}
$$
and one similar to the non-zero mode Feynman rule
in (\ref{modcorrelators})
$$
\langle \xi^\mu(\tau_1) \xi^\nu(\tau_2) \rangle = 
\frac{1}{2} \left( \sign(\tau_1 - \tau_2)
- 2 \frac{\tau_1 - \tau_2}{T} \right) 
g^{\mu\nu} 
\, .
$$

 
\section{A Comment on Worldline Supersymmetry}
\label{SUSYcomments}

It is of interest to consider the supersymmetry present in the coherent 
state formulation of the quantum mechanical path integral 
in the case where there is only an external vector field
(and no background axial or spinor fields).  These considerations
will demonstrate the importance of having anti-periodic boundary conditions
for the Grassmann variables being integrated.

It is known
that the supersymmetry properties of the worldsheet string action
put strong restrictions on the cycle structure of the worldline approach
to amplitude calculations.  In fact, the kinematic factors 
must vanish
when the Bosonic and Fermionic Green's functions both satisfy
periodic boundary conditions
\cite{fratse,BernKosower}.   

Here, we show how a simplified worldline supersymmetry 
can be used to arrive
at the same conclusion without resorting to any analysis based on the
cyclic structure of the Green's functions.
  
We begin with  the following path integral expression
\begin{equation}
\label{B1}
K_{\pm} = \int \D x(\tau) \int \D \psi(\tau) e^{- \int^T_0 L(\tau) d\tau }
\end{equation}
where $L$ is given in equation (\ref{defL}) (with $A_5 = 0$),
$x(\tau)$ is a closed Bosonic path,
and $\psi(\tau)$ is a Fermionic trajectory with 
the boundary condition $\psi(T) = \pm \psi(0)$.

The one-loop effective action for the vector field $A$
is well-known to be related to $K_-$, but
if, instead we consider the path integral $K_+$ with {\em periodic}
boundary conditions on the $\psi(\tau)$ variable, then 
we notice that it possesses the following
supersymmetry
\begin{eqnarray}
\label{SUSY}
Q x^\mu & = & - 2 \psi^\mu \\ 
Q \psi^\mu & = &  {\dot x}^\mu \, .\nonumber
\end{eqnarray}
It is trivial to show that the supersymmetry generator
has the property 
\begin{equation}
\label{BRST}
Q^2 = - 2 \frac{d}{d \tau} \, .
\end{equation}

We now observe that
the Lagrangian $L$ is in the image of the supersymmetry generator, 
specifically,
\begin{equation} 
L = 
Q \left[ \frac{1}{4} \dot x \cdot \psi + i \psi \cdot A \right] 
\end{equation}
which, by (\ref{BRST}) is sufficient to show that 
$\int^T_0 d\tau L$ is supersymmetric as long as $x(\tau)$ and $\psi(\tau)$ 
are
both periodic.  We define $\Omega = \int^T_0 d\tau \left(
 \frac{1}{4} \dot x \cdot \psi + i \psi \cdot A \right) $.

We will now demonstrate, using the techniques of 
\cite{Thompson}, that
a supersymmetric Ward identity guarantees that $K_+$ is 
independent of the vector field $A$.
 
The functional derivative of $K_+$ with respect to $A^\mu(x)$ is
\begin{equation}
\label{omegamu}
\frac{\delta}{\delta A^\mu(x)} K_+
= - \int \D x(\tau) \int D\psi(\tau) e^{- Q \Omega }
Q \Omega_\mu(x)
\end{equation}
where $\Omega_\mu(x) = \frac{\delta \Omega}{\delta A^\mu(x)}$.

Using the standard path-integral derivation of Ward identities,
we now consider the path integral
\begin{equation}
\label{SUSYtest}
 \int \D x(\tau) \int \D \psi(\tau) e^{- Q \Omega }
\Omega_\mu(x)
\end{equation}
which should be invariant under the following change of variables
generated by (\ref{SUSY}) 
\begin{eqnarray}
\label{SUSYtrans}
x'^\mu & = & x^\mu + \epsilon Q x^\mu \nonumber \\
\psi'^\mu & = & \psi^\mu + \epsilon Q \psi^\mu  \\
\Omega' & = & \Omega + \epsilon Q \Omega \, \nonumber
\end{eqnarray}
where $\epsilon$ is some Grassmann number. 
{}From (\ref{SUSY}) it
is straightforward to see that this transformation
preserves the periodic boundary conditions of $x(\tau)$ and $\psi(\tau)$ 
and that
its Jacobian superdeterminant  
is unity.
The action $Q\Omega$ is invariant under
(\ref{SUSYtrans}).  Hence, we discover that
(\ref{SUSYtest}) is equal to
\begin{equation}
\label{SUSYtest2}
 \int \D x(\tau) \int \D \psi(\tau) e^{- Q \Omega }
\left[ \Omega_\mu(x) + \epsilon
Q \Omega_\mu(x) \right] \, .
\end{equation}
The equality of (\ref{SUSYtest}) and (\ref{SUSYtest2}) immediately
implies that (\ref{omegamu}) vanishes and that $K_+$ is 
perturbatively independent
of $A^\mu(x)$.

Considering the simplicity of this proof, the topological
nature of these amplitudes is surprisingly nontrivial to
see at the parameter integral level. Things work out neatly
for the case where only a constant background field is
present. Treating this case in Fock-Schwinger gauge along the
lines of \cite{ss1} one finds that the path integral determinant
factors into two Euler-Heisenberg
type determinants from $\int {\cal D}y$ and $\int {\cal D}\xi$,
which cancel by supersymmetry, and
a Grassmann zero mode contribution. 
This zero mode contribution is given, for $D=4n$, by the
$n$ - th Pontrijagin density, thus leading to a topological
number indeed (which vanishes for the abelian case).
However things are less transparent for the full $N$ - point amplitude.
Here the main term, not involving the Grassmann zero mode,
can be seen to vanish after applying the 
partial integration procedure and the appropriate Bern-Kosower
substitution rule \cite{menphoton} (which originally was
derived in the string-context from worldsheet supersymmetry
\cite{BernKosower}). However there exist many more contributions
involving the zero mode, and the mechanism of their
vanishing seems not obvious.

\eject
\section{Second-Order Feynman Rules}
\label{secondorderrules}

Here we give the second-order (Euclidean) 
Feynman rules for the Lagrangian
(\ref{e100}), as following from eq.(\ref{e2}). 
We have now restituted the coupling constants,
$A\rightarrow eA, A_5\rightarrow e_5 A_5$.
The photon
propagator is the usual one.

\newlength{\fdwidth}
\setlength{\fdwidth}{1in}

\newlength{\frwidth}
\setlength{\frwidth}{1.0in}
\begin{figure}[h]
\centering
\begin{tabular}{r@{\hspace{0.2in}}l@{\hspace{0.6in}}r@{\hspace{0.2in}}l}
\raisebox{0.2in}{\epsfig{file=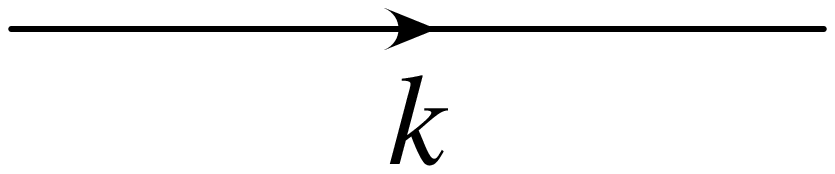, width=\fdwidth} } 
& \raisebox{0.65in}{\parbox{\frwidth}{\begin{displaymath} \frac{1}{k^2
+m^2} 
\end{displaymath}}} 
& \epsfig{file=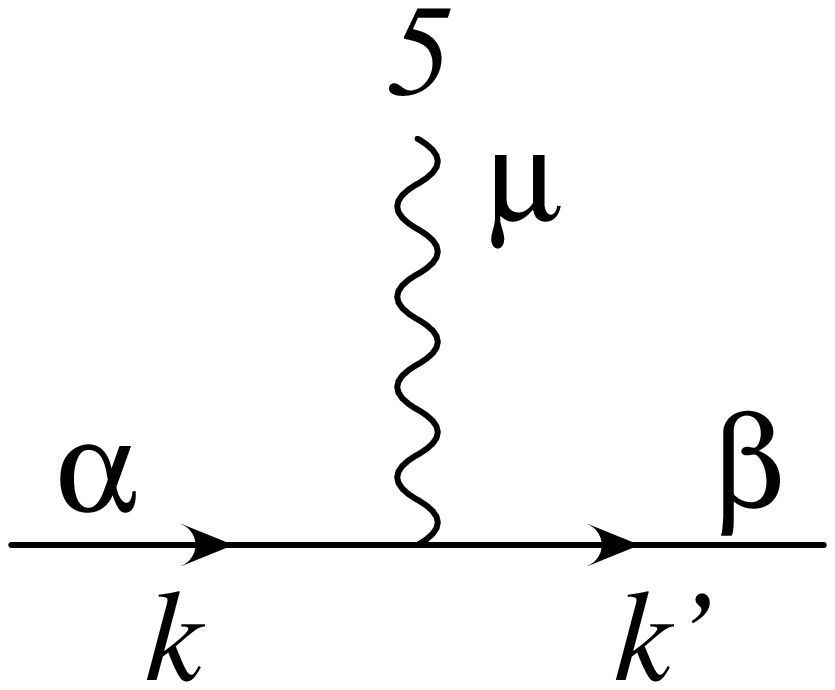, width=\fdwidth} 
& \raisebox{0.65in}{\parbox{\frwidth}{$$e_5 
(\gamma_5 \sigma_{\mu\nu})_{\beta\alpha}
\left( k + k' \right) ^ \nu $$ }}
\\
\epsfig{file=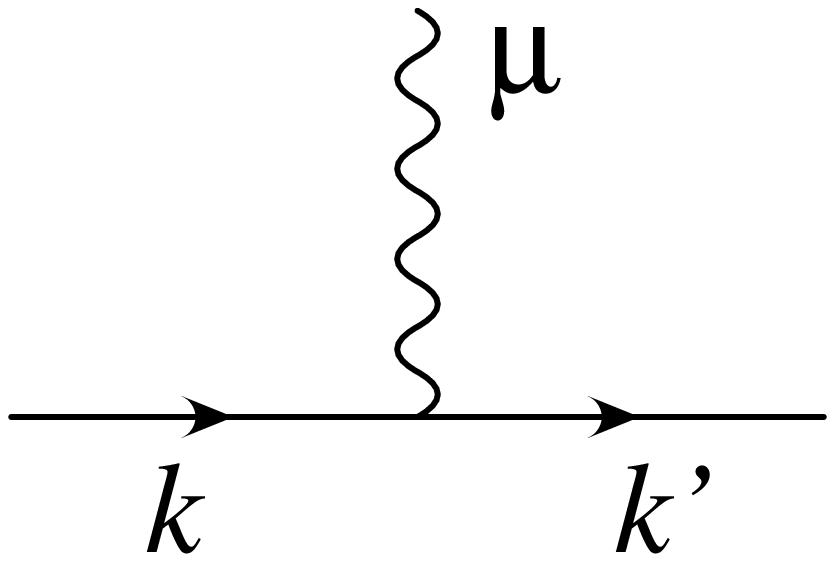, width=\fdwidth}   
& \raisebox{0.6in}{\parbox{\frwidth}{$$ e \left( k + k'\right)_\mu $$}} 
& \epsfig{file=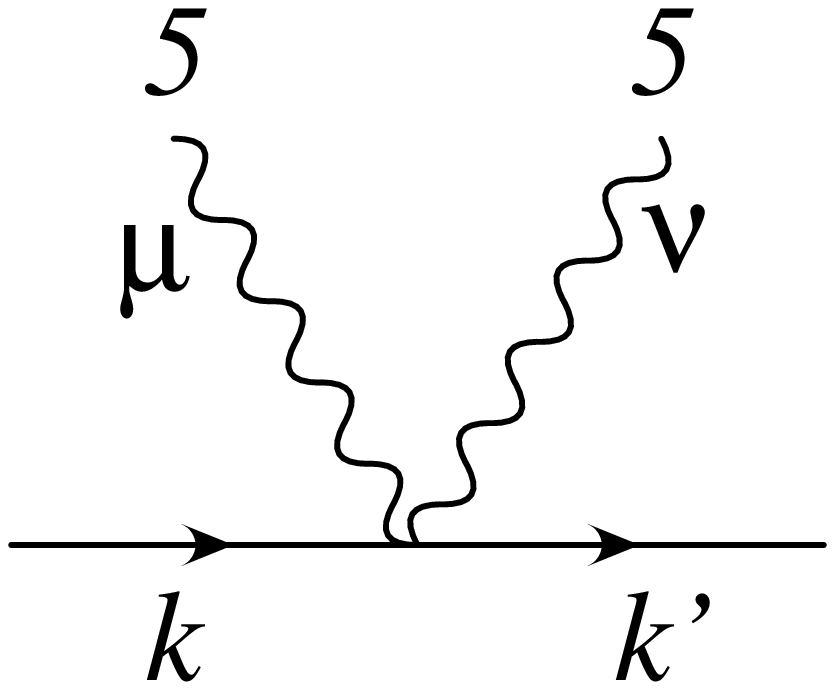,width=\fdwidth} 
& \raisebox{0.6in}{\parbox{\frwidth}{$$ 2 e_5^2 g_{\mu\nu}$$}}
\\
\epsfig{file=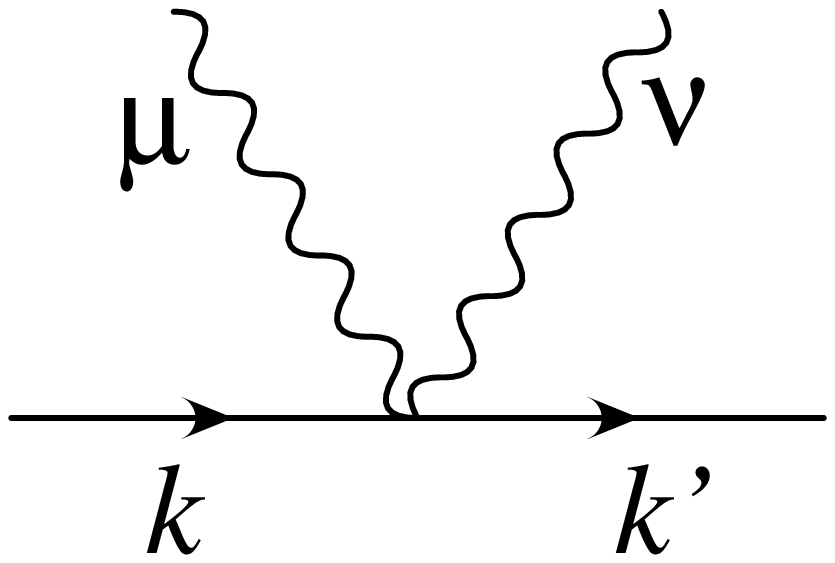,width=\fdwidth} 
& \raisebox{0.6in}{ \parbox{\frwidth}{ $$-2e^2g_{\mu\nu}$$ } }
& \epsfig{file=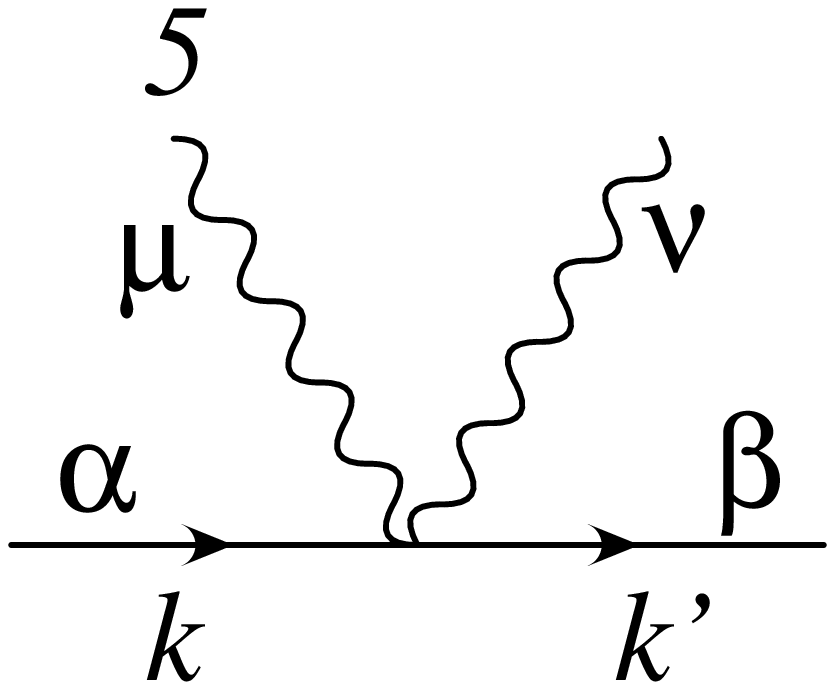,width=\fdwidth}
& \raisebox{0.6in}{\parbox{\frwidth}{$$-2 e e_5 (\gamma_5
\sigma_{\mu\nu}
)_{\beta\alpha}$$}}\\
\epsfig{file=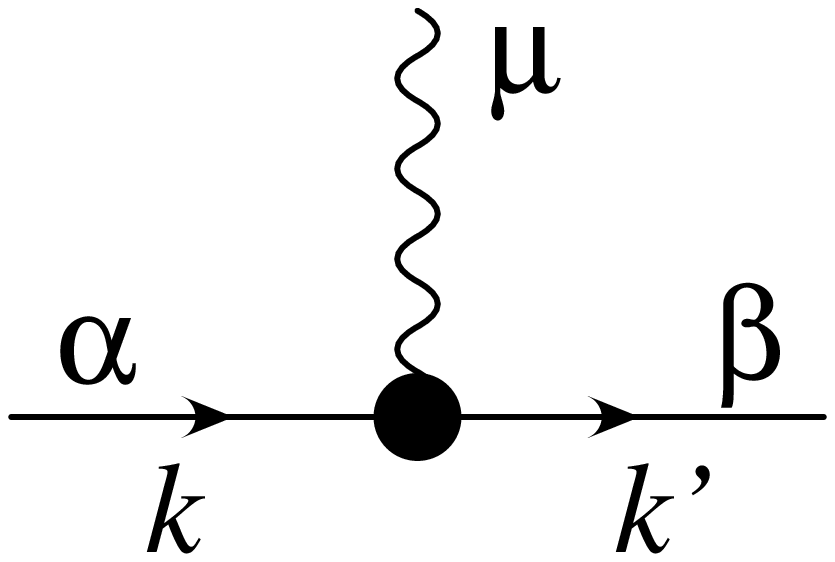,width=\fdwidth} 
& \raisebox{0.6in}{\parbox{\frwidth}{$$e (\sigma_{\mu\nu})_{\beta\alpha}\left
(k'-k\right )^\nu$$} }
& \epsfig{file=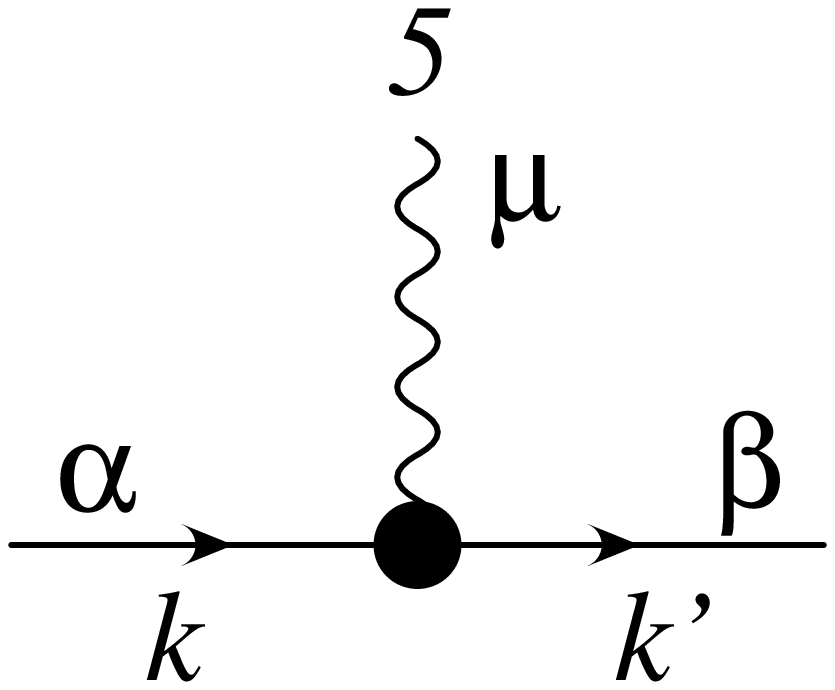,width=\fdwidth} 
& \raisebox{0.6in}{\parbox{\frwidth}{$$ e_5 
(\gamma_5)_{\beta\alpha}\left( k' - k \right)_\mu$$}}
\end{tabular}
\caption{Second order Feynman rules}
\end{figure}
\no
A factor $-\half$ is to be supplied for a closed fermion loop,
as well as a spinor trace (even for diagrams not involving Dirac
matrices, which then carry the unit matrix). For more information on 
the second order formalism see \cite{morgan}. Note that those
rules remain valid in dimensional regularization with no further
changes; there is no analogue of the $D$ - dependent coefficient
which we had in the worldline Lagrangian.

The calculation of the one-loop amplitude for $M$ vectors and
$N$ axialvectors with these rules yields, using Feynman parameters,
precisely the same integrands as we have in our master formulas
eqs. (\ref{axialevenmaster}),(\ref{axialoddmaster}). However the
master formulas already give the complete amplitude, with all
crossed diagrams included. An individual diagram is represented
by an ordered sector $\tau_{i_1}\geq \tau_{i_2}\geq \cdots$, the
Feynman parameters being the differences of adjacent $\tau_{i_j}$ 
- parameters
(with our conventions the charge flow is in the direction of
{\sl ascending} $\tau_{i_j}$).

\end{document}